\documentclass[aps,prc,twocolumn]{revtex4}
\usepackage{xcolor}
\usepackage{graphicx}
\usepackage{amsmath,bbm}
\usepackage{amssymb,bm}
\usepackage{array}
\usepackage{enumerate}
\usepackage{slashed}
\newcommand\nd{\noindent}
\newcommand \sH {\mathcal{H}}
\newcommand \sL {\mathcal{L}}

\newcommand \sN {\mathcal{N}}
\newcommand \bx {{\bf{x} }}
\newcommand \bp {{\bf{p} }}
\newcommand \bq {{\bf{q} }}
\newcommand \dV {{ \mathbb{V} }}

\DeclareMathOperator{\cosech}{cosech}
\begin{document}
\title{ Quantum theory, thermal gradients and the curved Euclidean space}
\author{S.~Ganesh\footnote{Corresponding author:\\Email: gans.phy@gmail.com}}
\affiliation{Sri Sathya Sai Institute of Higher Learning, Prasanthi Nilayam - 515134, A.P., India}

\begin{abstract} 
	The Euclidean space, obtained by the analytical continuation of time, to an imaginary time, is used to model thermal systems.
        In this work, it is taken a step further to systems with spatial thermal variation, by developing an equivalence between the spatial variation of temperature in a thermal bath and the curvature of the Euclidean space. The variation in temperature is recast as a variation in the metric, leading to a curved Euclidean space.
	The equivalence is substantiated by analyzing the Polyakov loop, the partition function and the periodicity of the correlation function. 
	The bulk thermodynamic properties like the energy, entropy and the Helmholtz free energy are calculated from the partition function, for small metric perturbations, for a neutral scalar field. 
The Dirac equation for an external Dirac spinor, traversing in a thermal bath with spatial thermal gradients, is solved in the curved Euclidean space. 
The fundamental behavior exhibited by the Dirac spinor eigenstate,
	may provide a possible mechanism to validate the theory, at a more basal level, than examining only bulk thermodynamic properties.
	Furthermore, in order to verify the equivalence at the level of classical mechanics, the geodesic equation is analyzed in a classical backdrop.
The mathematical apparatus is borrowed from the physics of quantum theory in a gravity-induced space-time curvature.
        As spatial thermal variations are obtainable at QCD or QED energies, it may be feasible for the proposed formulation to be validated experimentally.
\vskip 0.5cm

	{\nd \it Keywords } : Quantum theory, Curved Euclidean space, Dirac Equation, Thermal gradient, Equivalence, Scalar Field \\
{\nd \it PACS numbers } : 04.62.+v, 11.10.Wx, 03.65.-w

\end{abstract}

\maketitle

\section{Introduction}
\label{sec:intro}
   Systems with spatial thermal gradients is ubiquitous from the Quantum Chromodynamic (QCD) systems like the Quark Gluon Plasma (QGP)~\cite{nature,naturephy}, to astronomical objects like the interior of the Sun. This paper focuses on the smaller scales where quantum mechanical effects can be influenced by thermal gradients.
  The Euclidean space-time, obtained by analytical continuation of time in Minkowski space-time to an imaginary time, is often used to model thermal systems~\cite{matsubara, martin, gorkov, adas}.
More recent literature on the applications of thermal field theory is available in Refs.~\cite{misc1, misc2, Kraemmer, Kraemmer2, misc3, misc7, misc4, misc5, misc6, tft, epja}.
  The temperature is modeled as the inverse of imaginary time. 
  In this work, the temperature variation is modeled by recasting it as a variation in the metric. This leads to a curved Euclidean space.
  In Ref.~\cite{gans5}, this concept has been used to evaluate the Wilson loop using gauge-gravity duality, for the purpose of calculating the effect of temperature gradient on quark, anti-quark potential. The current work places the concept on a firmer footing by exploring the important aspects of imaginary time formalism, like the partition function and the correlation function, in a curved Euclidean space. 
The thermal bath needs to be in local thermal equilibrium. The current formalism may not be good enough for systems in non-equilibrium or very far from equilibrium.
  The analysis is restricted to a canonical ensemble, and thus would be applicable to a system with vanishing chemical potential. 
  However, the grand canonical ensemble is touched upon in Sec.~\ref{sec:gce}.
  The thermal bath is considered to be static (time independent). 
This work also elaborates the physical implications of the theory which can be used to validate the theory experimentally.

A covariant approach could involve considering a complex 4-D space as $X^{\mu} \equiv x^{\mu} + i\beta^{\mu}$, with $x^{\mu}$ being the conventional space-time, and $\beta^{\mu} = \beta u^{\mu}$, where, $u^{\mu}$ is the four-velocity representing the flow of the thermal medium, and $\beta$ is the inverse temperature. 
It is trivially seen that this space, under a Lorentz boost, would transform as, $X'^{\nu} = \Lambda^{\nu}_{\mu} X^{\mu}$, where $\Lambda^{\nu}_{\mu}$ represents the Lorentz transformation.
If the thermal medium is stationary, i.e., $u^{\mu} = (1,0,0,0)$, then the complex space-time reduces to $(t + i\beta, x, y, z)$. 
In the case of a time invariant system, it is sufficient to consider the sub-space $(i\beta, x, y, z)$, which forms the co-ordinate space for the imaginary time formalism, considered in this work. 
One might also be tempted to view the original 4-D complex space, $x^{\mu} + i\beta^{\mu}$, as a 8-D, $\beta^{\mu}\times x^{\nu}$, space. 
The 8-D space under a Lorentz boost would transform as: 
   		\begin{equation}
		\label{eq:8D}
		\left[ \begin{array}{c}
			\beta'^{\gamma}\\
			x'^{\delta}\\
		\end{array} \right ] 
			 = 
		\left[ \begin{array}{c c}
			\Lambda^{\gamma}_{\alpha}&0\\
			0&\Lambda^{\delta}_{\beta}\\
		\end{array} \right ] 
		\left[ \begin{array}{c}
			\beta^{\alpha}\\
			x^{\beta}\\
		\end{array} \right ]. 
   		\end{equation}
		The sub-space, $\beta^{\mu}$, can be thought of as the sub-space spanned by the thermal bath. The sub-space, $x^{\nu}$, could be used to describe the system immersed in the thermal bath. Additionally, $\beta^{\mu}$ could itself depend on $x^{\nu}$.
A formulation that captures the interaction between the thermal bath and the system within it, would be expected to involve both the sub-spaces.
Let us consider the following metric tensor in this 8D space:
   		\begin{equation}
			\label{eq:8Dmetric}
			G^{8D} = 
		\left[ \begin{array}{c c}
			h_{\rho \sigma}&0\\
			0&g_{\mu\nu}\\
		\end{array} \right ].
   		\end{equation}
The $\rho,\sigma,\mu$ and $\nu$ indices vary within their 4-D subspaces respectively.
The metric elements, $g_{\mu\nu}$, are the usual metric elements describing the curvature of space-time, $x^{\mu}$, under a gravitational field. The interpretation of $h_{\rho\sigma}$ is yet to be ascertained.
In the limit $u^{\mu} = (1,0,0,0)$, 
the 8-D $\beta^{\mu} \times x^{\nu}$ space, 
reduces to the five dimensional space $(i\beta, t, x, y, z)$. 
Under the additional more restrictive conditions,
\begin{itemize}
\item the thermal bath and the system immersed in it is time invariant,
\item  $g_{00} = 1$; $g_{0i} = 0~\forall i = 1,2,3$,
\end{itemize}
it suffices to consider the 4-D Euclidean subspace, $(i\beta,x,y,z)$, which again leads to the imaginary time formalism.
The metric for the 4-D sub-space, $(i\beta, x, y, z)$, can be inferred from Eq.~\ref{eq:8Dmetric} as:
   		\begin{equation}
			\label{eq:4Deuc_metric}
			G^{Euc} = 
		\left[ \begin{array}{c c}
			h_{00}&0\\
			0&g_{ij}\\
		\end{array} \right ],
   		\end{equation}
where $i,j = 1,2,3$.
The rest of the paper develops the interpretation of $h_{00}(x,y,z)$ as the spatial thermal gradient. In other words, the equivalence between a curved Euclidean space and spatial thermal gradients is developed. 
It may be emphasized, that, treating $t$ and $i\beta$ as distinct dimensions, provides a simple mechanism to enable the metric elements, $g_{00}$ and $h_{00}$, to be independent of each other.
In particular, the Euclidean space is curved, whilst, the underlying Minkowski space-time is flat. 
Another consequence of the 8-D space is the doubling of the degrees of freedom, which is shown as a necessity to define a thermal vacuum in thermo field dynamics~\cite{tft1, tft2}. 
Since, the current work deals with the special case of a 4D Euclidean subspace, it can be seen as a step towards a generic covariant treatment of systems with both spatial and temporal thermal gradients.
For the sake of future reference, the gravity induced metric in the Lorentzian space-time is denoted as:
\begin{equation}
\label{eq:4Dgrav_metric}
	G^{grav}_{\mu\nu} = g_{\mu\nu},
\end{equation}
with $\mu,\nu = 0,1,2,3$.
The values of $g_{ij} (i,j = 1,2,3)$ are actually irrelevant to the development of the proposed formalism. Since $g_{00}$ is already taken as 1 in the context of imaginary time formalism, one can then assume $g_{\mu\nu} = \eta_{\mu\nu}$, without any further loss of generality.

  In his seminal work, Matsubara~\cite{matsubara}, had evaluated the partition function using the mathematical apparatus of quantum field theory (QFT).
  In the current work, we show that the evaluation of the partition function, in the context of a temperature gradient, is consistent with the reformulation of a system with a temperature gradient as a curved Euclidean space. 
  The correlation function is central to the calculation of the propagators in the imaginary time formalism. The periodicity of the correlation function, leads to a cylindrical topological space. It is shown that the periodicity of the correlation function is maintained when the temperature variation is recast as a variation in the Euclidean metric.
  After the above theoretical foundations, the partition function is used to calculate bulk thermodynamic parameters like energy, Helmholtz free energy and entropy, for a neutral scalar field in the curved Euclidean space.
  A curved Euclidean space leads to the introduction of geodesics. The meaning and interpretation of the geodesic in the Euclidean space is explored.
  Comparing the geodesic equations with the kinetics of a test particle in a classical framework, helps to explore the equivalence at the level of classical mechanics. 
  These aspects help to establish the concept of a curved Euclidean space to model temperature gradients, on a firmer footing.
	The equivalence between the spatial thermal gradients and the curved Euclidean space is introduced using a Wilson-Polyakov loop. This has been explained in Ref.~\cite{gans5}, but is mentioned here again for clarity. The equivalence is  then substantiated by developing the partition function, periodicity of the correlation function and the geodesic equation.

Having established a reasonable justification for such a mathematical model to capture thermal gradients, the question arises, whether there is any non-trivial, peculiar physical phenomenon, which can be used to check if the hypothesis is or is not a valid description of the physical world.
The thermodynamic properties of a neutral scalar field, and the interpretation of the geodesic equation, at a classical level, are steps towards this objective. 
For a more persuasive phenomenon, the Dirac equation is subsequently explored in the curved static Euclidean space, in Sec.~\ref{sec:dirac}. This involves modeling an external Dirac spinor, traversing a thermal system with spatial thermal gradients.
The fundamental behavior of the eigenstate is intriguing.
One would see later in Fig.~\ref{fig:l1}, that, even when the temperature is non-zero, in the case of a negative temperature gradient, the wave-function can actually increase in a thermal bath. This aspect seems peculiar, as one would normally expect a wave-function to only decay in a thermal bath. If this is observed experimentally, it can provide evidence that the curved Euclidean space approach is indeed a valid mathematical description of the physical systems having a temperature gradient.
The analysis of the behavior of a single eigenstate instead of the bulk thermodynamic properties, might be a more fundamental validation of the proposed theory.
While writing this manuscript, it has come to my attention, that, Ref.~\cite{hongo} has proposed an analogous formalism for modeling local thermal equilibrium in the context of a hydrodynamic flow. 
Reference~\cite{hongo}, has attempted to model the thermal bath in a covariant way, by considering the "space-time" to be spanned by $\beta^{\mu} = \beta u^{\mu}$, where $u^{\mu}$ is the four-velocity of the thermal bath. 
However, for a static thermal bath, with $u^{\mu} = (1,0,0,0)$, the space-time reduces to a 1-D space. 
Additionally, it considers only the 4-D space, $\beta^{\mu}$, instead of the 8-D space, $\beta^{\mu} \times x^{\nu}$.
Consequently, the formalism in Ref.~\cite{hongo}, could preclude certain important phenomenon of interest, like the one highlighted in this paper, in Sec.~\ref{sec:dirac}, from being manifested. 
Furthermore, the 4D sub-space of $\beta^{\mu}\times x^{\nu}$, i.e., the sub-space $(i\beta,x,y,z)$, is required for the equivalence hypothesis in Sec.~\ref{sec:polyakov}.
Thus, there are fundamental differences between the current approach and the approach adopted in  Ref.~\cite{hongo}.

If the curved Euclidean space hypothesis indeed describes the physical world, it may open up interesting possibilities.
Since the energies required for obtaining a curved Euclidean space can be obtained in the lab, it may be feasible to validate the quantum theory in a curved Euclidean space experimentally, unlike the quantum theory in a gravitational field. 

  The primary goal of this paper is to help establish the concept of a curved Euclidean space, i.e., $G^{Euc} \ne I$, for modeling thermal gradients. However, it could be tempting to compare it with quantum mechanics in a curved Lorentzian space, i.e., $G^{grav}_{\mu\nu} \ne \eta_{\mu\nu}$, as it could provide insights for either theory.
  Quantum theory, in a gravity induced curved space-time, with Lorentzian signature (referred to as a curved Lorentzian space for short), is a topic of intense research~\cite{book1,mukh,padh1,padh2,arb1,c1,c2,c3}. 
  In fact, it has led to many interesting phenomena like the Hawking radiation~\cite{hawking} and Unruh effect~\cite{unruh}. However, testing the quantum theory in a curved space-time is a formidable challenge due to the high energies required to curve space-time. 
  It may be interesting to analyze the similarities and differences between the quantum theory in the curved Lorentzian space and the quantum theory in the curved Euclidean space.
  We elucidate the effect of the curvature on a fermion wave-function, when it passes through the curved Euclidean space ($G^{Euc} \ne I$), and a curved Lorentzian space ($G^{grav}_{\mu\nu} \ne \eta_{\mu\nu}$), by solving the Dirac equation, in Sec.~\ref{sec:diracgravity}. 
  The effects are not identical, and the 
  the key would be the identification of the mathematical similarities and differences, between the two curved spaces.
  In the case of mathematical similarities, it might be possible that experimental exploration of a quantum theory in a curved Euclidean space, may provide limited insights into the quantum theory in a gravity induced curved Lorentzian space-time.

Another phenomenon that bears mathematical resemblance to the proposed phenomenon, but occurs in the curved Lorentzian space, is the Ehrenfest Tolman effect~\cite{ET1,ET2}.
The Ehrenfest Tolman phenomenon is about the appearance of thermal gradients, in a thermal system at global equilibrium, in the presence of a gravitational field. 
This topic is touched upon in Sec.~\ref{sec:partition}.


  The  thermal bath can either be a QCD plasma or a Quantum Electrodynamic (QED) plasma~\cite{plasma3,plasma2}.
 If the thermal bath were to be a QCD plasma, namely the QGP, then the energies involved would be in the GeV scale, and the distances involved would be in femtometer scale. 
  QGP is currently being produced at the Large Hadron Collider (LHC) and the Relativistic Heavy Ion Collider (RHIC)~\cite{nature}.
 If, however the thermal bath were to be a QED plasma, then the energies reduce further to keV or even tens of ev~\cite{plasma1}, and the distances involved would be in nanometer (nm) or picometer (pm) range.
 Additionally, it may be easier to obtain the stationarity (not time dependent) conditions, required for the applicability of the proposed formalism, in a QED plasma. 

 The rest of the paper is as follows. The equivalence is first intuitively introduced using a Wilson Polyakov loop correlator in Sec.~\ref{sec:polyakov}. 
 The Euclidean path integral formulation of the partition function is then developed for a system in local thermal equilibrium in Sec.~\ref{sec:partition}, which substantiates the mathematical equivalence with a curved Euclidean space. 
 Section~\ref{sec:periodicity}, explores the periodicity of the correlation function.
 For small variations in the temperature, the thermodynamic parameters like energy, entropy and the Helmholtz free energy are calculated for a neutral scalar field in Sec.~\ref{sec:scalarfield}.
    The geodesic equation, in a classical context, is explored in Sec.~\ref{sec:geodesic}.
    In Sec.~\ref{sec:dirac}, we solve the Dirac equation, which models an external massive Dirac spinor, traversing a curved Euclidean space. This section also captures the numerical work performed for solving the Dirac equation. The effect of spatial curvature on the fermion 3-momentum is explored.
  This is then compared with the Dirac equation in a curved Lorentzian space-time.
The physical interpretations of aspects like the Matsubara eigenstates and other subtle aspects of the formalism are covered in Sec.~\ref{sec:interp}.
  Finally, in Sec.~\ref{sec:conclusion}, we draw the conclusions.

\section{The Equivalence: A Hypothesis}
\label{sec:polyakov}
\begin{figure}
\includegraphics[width = 80mm,height = 80mm]{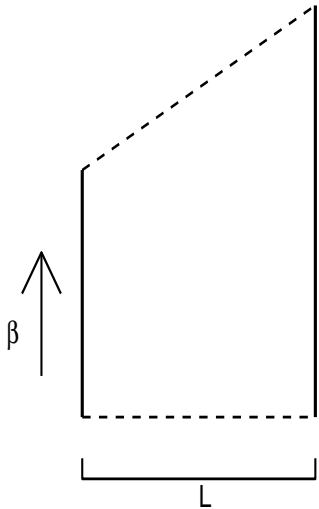}
\caption{Wilson-Polyakov loop to model the potential between two particles in a potential gradient.
	The two solid vertical lines represent the two Wilson lines, corresponding to the two Polyakov loops $P(0)$ and $P(L)$.}
\label{fig:single_loop}
\end{figure}

Let us first consider a scenario, where a heavy Quark and anti-Quark are placed in a thermal bath, having a temperature gradient.
The Wilson-Polyakov loop that has been used to model the scenario of the temperature gradient is depicted in Fig.~\ref{fig:single_loop}~\cite{gans5}. 
The relation of the Wilson-Polyakov loop correlator to the potential between the Quark and anti-Quark could be modeled as~\cite{soo}:   
\begin{equation}
\label{eq:qqbeta}
<P(0)P(L)> = e^{-V_{Q\bar{Q}}\beta},
\end{equation}
where, $\beta = \frac{1}{\Theta}$ 
and $\Theta$ is the temperature of the system. $V_{Q\bar{Q}}$ is the potential between the Quark and anti-Quark. In the case of a QCD system, the two particles in the system are the Quark and anti-Quark. One can however, choose a QED system, in which the particles in the thermal bath could be  two oppositely charged leptons, e.g., muon and anti-muon. The Polyakov loop is the trace of the Wilson line integrated from 0 to $\beta$ in the Euclideanized time, i.e., $P(x) = \frac{1}{N}Tr\Big [T \exp(-i\int_0^{\beta}A_0(x,\tau)d\tau \Big ]$. Here, $A(x,\tau)$ is the pure gauge field. In the case of QCD, a time ordering operator, $T$, would be required. 
While evaluating the Polyakov loop correlator, the Minkowski space time is converted to a Euclideanized time by the use of an imaginary periodic time.

If there is a temperature gradient, then the R.H.S. of Eq.~\ref{eq:qqbeta}, poses a problem. There is no provision to have a spatial variation of $\beta$. To resolve this issue, let us probe the Euclidean space further.
In the imaginary time formalism, the space-time fabric is a uniform topological cylinder, in a Euclidean space of radius $\frac{\beta}{2\pi}$. The contour in Fig.~\ref{fig:single_loop}, seems to indicate that the cylinder is non-uniform, with radii $\frac{\beta}{2\pi}$ and $\frac{\beta + \Delta \beta}{2\pi}$ at the two ends. Thus, application of Eq.~\ref{eq:qqbeta} becomes a challenge. However, it is possible to alternatively view the topological cylinder as a uniform cylinder in a curved Euclidean space (a straight line in curved space would be curved). 
 If one were to evaluate the Polyakov loop correlator, in Eq.~\ref{eq:qqbeta}, in the field theoretic domain, one would then in principle, evaluate the correlator in the curved Euclidean space, with a "constant" $\beta$. 
In Sec.~\ref{sec:periodicity} and \ref{sec:dirac}, we determine the value of this constant $\beta$.

 Another perspective is that, in a Minkowski space-time, if time were to depend on the spatial location, then, the dependence can be seen as a result of curvature in the Minkowski space-time. Analogously, if the temperature were to depend on spatial location, then the Euclidean space should be curved. 

\begin{figure}
\includegraphics[width = 70mm,height = 70mm]{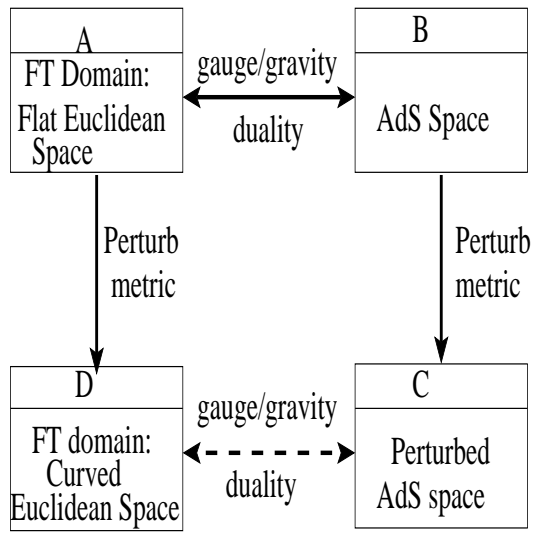}
\caption{"FT Domain" stands for "Field Theoretic Domain". Block $C$ and block $D$ would be dual of each other due to the gauge-gravity duality.}
\label{fig:perturb}
\end{figure}

 This gives rise to an equivalence principle, that variation in temperature can be recast as a variation in the metric, leading to a curved Euclidean space.
This equivalence principle has been used in Ref.~\cite{gans5} to evaluate the quark anti-quark potential in the presence of thermal gradients, using the gauge-gravity duality. 
An inverse argument can also be made. In the gravity dual domain, the temperature gradient is introduced by perturbing a black string metric, which in turn perturbs the AdS space.
Due to gauge-gravity duality, the perturbation of the AdS space, translates to the perturbation of the Euclidean space in the field theoretic domain.
A diagrammatic representation of how the gauge-gravity duality naturally leads to the equivalence principle, is depicted in Fig.~\ref{fig:perturb}.
Figure~\ref{fig:perturb}, summarizes the basic concept behind Ref.~\cite{gans5}.
In Fig.~\ref{fig:perturb}, block $C$ is obtained by perturbing the AdS space (block $B$). The Euclidean metric in block $A$, is equivalently perturbed to obtain block $D$. Gauge-gravity duality, then implies that block $D$ and block $C$ are duals of each other.
Thus, the calculation of the Wilson loop in Fig.~\ref{fig:single_loop} using gauge-gravity duality, in Ref.~\cite{gans5}, can be seen to substantiate the proposed equivalence hypothesis.

 Since spatial temperature variation is required, the thermal bath (either the QCD or QED thermal bath), would be in a state of local thermal equilibrium, instead of a full-fledged global thermal equilibrium.

Before concluding this section, let us explore the mechanism of arriving at a desired spatial temperature profile, which is stationary in time.
From Ref.~\cite{book3}, the variation of temperature with time is given by:
\begin{equation}
	\label{eq:heateq}
	\rho C_v \frac{\partial \Theta}{\partial t} = {\dot q} + \nabla.k\nabla\Theta,
\end{equation}
where $\rho$ is the density, $C_v$ is the specific heat at constant 
volume, $k$ is a constant and $q$ is the energy generated (or sinked) per unit volume.
For temperature to be stationary in time ($\frac{\partial \Theta}{\partial t}=0$), we get,
\begin{equation}
	\label{eq:zerotempgrad}
	{\dot q} = -\nabla.k\nabla\Theta.
\end{equation}
Thus, with the help of appropriate external heat sources and sinks, a desired temperature profile could be generated, which is stationary in time.

\section{The Partition Function}
\label{sec:partition}
The partition function is a very fundamental thermodynamic quantity.

A time invariant Hamiltonian lends itself to a more explicit analysis. Hence we shall start with a time invariant Hamiltonian,
and subsequently, touch upon the case of a generic Hamiltonian. In either case, the temperature does not vary with time, i.e. $\frac{\partial \Theta}{\partial t} = 0$.
\subsection{Time invariant Hamiltonian}
\label{sec:timeinvariant}
Kapusta and Gale~\cite{kapusta}, have outlined the derivation of the partition function in a functional integral form for a time invariant Hamiltonian.  The derivation in this section parallels Ref.~\cite{kapusta}, albeit, modified for spatial thermal gradients. 
Let, 
\begin{eqnarray}
\nonumber	\hat{\phi}(\bx,0)|\phi\rangle = \phi(\bx)|\phi\rangle,\\
	\hat{\pi}(\bx,0)|\pi\rangle = \pi(\bx)|\pi\rangle.
\end{eqnarray}
In other words,	 $\phi(\bx)$ and $|\phi(\bx)\rangle$ are the eigenfunction and the eigenket of the Schrodinger picture field operator $\hat{\phi}(\bx,0)$, while, $\pi(\bx)$ and $|\pi(\bx)\rangle$ are the eigenfunction and the eigenket of the  conjugate momenta field operator $\hat{\pi}(\bx,0)$.

In the framework of the generalized Fourier transformation, we also have,
\begin{equation} 
	\langle \phi| \pi \rangle = \exp\left (i\int d^3x \pi(\bx)\phi(\bx) \right ).
\end{equation} 
We rewrite the Hamiltonian, $H$, in terms of the spatial Hamiltonian density $\sH$~\cite{kapusta}.
\begin{equation}
	H = \int d^3x \sH(\hat{\pi}(\bx,0), \hat{\phi}(\bx,0)) \equiv \int d^3x \sH(\bx).
\end{equation}
For economy of notation, the form $\sH(\bx)$ is used as a shortened representation of $\sH(\hat{\pi}(\bx,0), \hat{\phi}(\bx,0))$. 
   
The evolution of a quantum state from $|\phi_a\rangle$ to a state $|\phi_b\rangle$, in time $t$, is given by
\begin{math}
\nonumber	 \langle \phi_b| U(H,t)|\phi_a\rangle, 
\end{math}
with,
\begin{math}
\nonumber	U(H,t) = e^{-i\int Hdt}.
\end{math}
The analytical continuation, $t \rightarrow -i\tau$, gives,
\begin{math}
	\nonumber	U(H,\beta) =  \exp \left ( \int_0^\beta -Hd\tau \right ).
\end{math}
We wish to extend this for systems with spatial thermal variation, $\beta(\bx)$.
We represent $\beta(\bx) = s(\bx)\beta_0$, with $\beta_0 = \beta(\bx =(0,0,0))$, i.e., $s(\bf{0}) = 1$.
This relegates the spatial variation to $s(\bx)$.
Then,
\begin{eqnarray}
\label{eq:uhb}
        \nonumber U(\sH,\beta_0) =  \exp\left (-\int\int_0^{\beta(\bx)} \sH(\bx) d\tau d^3x\right ) \\
        =  \exp\left (-\int s(\bx)\sH(\bx)\beta_0 d^3x \right ).
\end{eqnarray}
To develop the path integral formulation, we proceed in a manner, which is very similar to the uniform temperature case.
Let $\beta_0$ be sliced into $N$ slices, i.e., $\beta_0 = N\Delta \beta$, with $N \rightarrow \infty$ and $\Delta \beta \rightarrow 0$. This gives,
\begin{eqnarray}
\nonumber	U(\sH,\beta_0)	  = \lim_{\Delta \beta \rightarrow 0} 
	\exp\left (-\sum\limits_{n} \left [ \int s(\bx)\sH(\bx) d^3x \right ] \Delta \beta  \right )
\end{eqnarray}
\begin{equation}
\label{eq:noneucaction}
	= \lim_{\Delta \beta,  \rightarrow 0} 
	  \prod\limits_{n}\exp\left (-\left [ \int s(\bx)\sH(\bx) d^3x \right ]\Delta \beta  \right ).
\end{equation}
Let $|\pi_j\rangle$ be the conjugate momentum eigenstate.
Inserting a complete set of eigenstates:
\begin{equation}
\nonumber I =  \int |\phi_j\rangle \langle \phi_j| d\phi_j \times \int |\pi_{j-1}\rangle \langle \pi_{j-1}| \frac{d\pi_{j-1}}{2\pi}, 
\end{equation}
between each product term in Eq.~\ref{eq:noneucaction}, and
\begin{equation}
	\nonumber I =   \int |\pi_{N-1}\rangle \langle \pi_{N-1}| \frac{d\pi_{N-1}}{2\pi};~~~
I =   \int |\phi_{1}\rangle \langle \phi_{1}| d\phi_1,
\end{equation}
in the beginning and the end of R.H.S. of Eq.~\ref{eq:noneucaction}, respectively, 
we get the expression for 
$\langle \phi_f|U(\sH,\beta_0)|\phi_0\rangle$, represented by $K(\phi_f,\phi_0,\beta_0)$:
\begin{eqnarray}
	\label{eq:pathintinter1}
\nonumber    K(\phi_f,\phi_0,\beta_0) = \lim_{\Delta \beta \rightarrow 0}  
	\int \prod\limits_{j=1}^{N-1} \langle \phi_{j+1}|\pi_j\rangle\\
\nonumber	\times \langle \pi_{j}| \exp\left (-\int s(\bx)\sH(\bx) \Delta \beta d^3x \right )|\phi_j\rangle \langle \phi_1| \phi_0 \rangle d\phi_j \frac{d\pi_j}{2\pi},\\
\end{eqnarray}
where, $\phi_N = \phi_f$.
$Z(\beta_0) = tr[K(\phi_f,\phi_0,\beta_0)]$ would provide the partition function in a thermal medium with spatial thermal distribution, $\frac{1}{\beta(\bx)}$. 
Equation~\ref{eq:pathintinter1}, contains the integrand $s(\bx)\sH(\bx)\Delta \beta d^3x$,
instead of $\sH(\bx)\Delta \beta d^3x $.
This can also be viewed as multiplying $\sH(x)$ with an infinitesimal 4-volume element
$\sqrt{|G^{Euc}|}d^3x\Delta \beta$ (as $\Delta \beta \rightarrow 0$), in a curved Euclidean space with metric $G^{Euc} = diag (s(\bx)^2,1,1,1)$. 
The factor, $s(\bx)^2$, is identified with the metric element $G^{Euc}_{00} = h_{00}(\bx)$ (refer Eq.~\ref{eq:4Deuc_metric}).
Thus, Eq.~\ref{eq:pathintinter1} can be interpreted as a 
path integral in a curved space with the metric, $G^{Euc}$.
In fact, the element, $s(\bx)^2$, can be identified with the metric element, $h_{00}(\bx)$, in Eq.~\ref{eq:uhb} itself, by re-writing the R.H.S. as $\exp \left (-\int \int_0^{\beta_0} s(\bx)\sH(\bx) d\tau d^3x \right )$.
Thus, either in the form of the path integral formulation, Eq.~\ref{eq:pathintinter1}, or the basic form in Eq.~\ref{eq:uhb}, the partition function naturally lends itself to the re-interpretation of a thermal gradient as a curvature of the Euclidean metric. 

We now evaluate Eq.~\ref{eq:pathintinter1}, and determine the partition function in the curved Euclidean space.
The expression in Eq.~\ref{eq:pathintinter1}, can be evaluated using the relations:
\begin{itemize}
\item
\begin{equation}
\label{eq:scalarpiphi}
	\langle \phi_{j+1}|\pi_j\rangle =  \exp \left (i\int d^3x \pi_j(\bx) \phi_{j+1}(\bx) \right ),
\end{equation}
\item
\begin{multline}
\label{eq:piexpphi}
\langle \pi_j| \exp \left (-i\int s(\bx)\sH d^3x \Delta \beta \right )|\phi_j \rangle   \\
	= \langle \pi_j|\phi_j \rangle \exp \left ( -i\int s(\bx)\sH_j d^3x \Delta \beta \right ), 
\end{multline}
with, $\sH_j = \sH(\pi_j,\phi_j)$,
\item and
\begin{equation}
\label{eq:phiphi}
	\langle \phi_1| \phi_0 \rangle = \prod_{\bx}\delta \left (\phi_1(\bx) - \phi_0(\bx) \right ).
\end{equation}
\end{itemize}

Inserting Eqs.~\ref{eq:scalarpiphi},~\ref{eq:piexpphi} and~\ref{eq:phiphi} in Eq.~\ref{eq:pathintinter1}, we obtain,
\begin{multline}
\label{eq:eucaction}
	K(\phi_f,\phi,\beta_0) = 
\lim_{\Delta \beta \rightarrow 0}  
	\int  \prod\limits_j d\phi_j \frac{d\pi_j}{2\pi} \exp \Bigg \{ \int d^3x\\
\times \Big [ i\pi_j(\bx)\Big \{ \phi_{j+1}(\bx) - \phi_j(\bx) \Big \} -  s(\bx)\sH_j  \Delta \beta \Big ] \Bigg \}
\end{multline}
\begin{multline}
	=\lim_{\Delta \beta \rightarrow 0} 
	  \int \left ( \prod\limits_j  d\phi_j \frac{d\pi_j}{2\pi} \right ) 
\exp \Bigg \{ \sum_l\Big (  \Delta \beta \int d^3x \\
	\times	s(\bx)  \left [ i \pi_l(\bx)\frac{\left \{ \phi_{l+1}(\bx) - \phi_l(\bx) \right \} }{s(\bx) \Delta \beta } -  \sH_l  \right ] \Big ) \Bigg \}.
\end{multline}
Let us recollect that, $s(\bx) = \sqrt{h_{00}(\bx)}$. Moreover, for the metric $G^{Euc} = (s(\bx)^2,1,1,1)$, we have, $\sqrt{|G^{Euc}|} = s(\bx)$. Denoting the determinant of the Euclidean metric, $|G^{EUC}|$, by $g_e$, we obtain in the continuum limit: 
\begin{eqnarray}
\label{eq:pathintfinal}
	\nonumber		   K(\phi_f,\phi,\beta_0) = 
	\int D\phi \int \frac{D\pi}{2\pi}
	\exp \Bigg \{ \int_0^{\beta_0} d\tau \int d^3x  \sqrt{g_e}\\
\nonumber	\times	\left [ i\pi(\bx,\tau)\frac{1}{\sqrt{h_{00}(\bx)}}D_{0} \phi(\bx,\tau)   -  \sH(\pi(\bx,\tau),\phi(\bx,\tau))  \right ]  \Bigg \},\\
\end{eqnarray}
where, $D_{0}$ is the covariant derivative w.r.t. $\tau$. For a scalar $\phi(\bx,\tau)$, $D_{0}\phi(\bx,\tau) = \frac{\partial \phi(\bx,\tau)}{\partial \tau}$. 
The exponent in Eq.~\ref{eq:pathintfinal}, is consistent with the action in a curved space if $\pi(\bx,\tau)$ is identified as the proper $\pi$, i.e., $\pi = \sqrt{h_{00}}\pi^0 = i\sqrt{h_{00}} \frac{\delta L}{\delta \partial_0 \phi}$.
The form of Eq.~\ref{eq:pathintfinal}, incorporates the case where the 3-D space is also curved, i.e., $\int d^3x \rightarrow \sqrt{|G^{Euc}_{ij}|}d^3x, \forall i,j=1.2.3$. 
In this case, $\sqrt{g_e}d\tau d^3x = s(x)\sqrt{|G^{Euc}_{ij}|} d\tau d^3x$.
For this work, however, we continue to consider, $G^{Euc}_{ij} = I_3$.
Compactly,
\begin{eqnarray}
	\label{eq:transamp}
	K(\phi_f,\phi,\beta_0)  
	=  \int  D\phi \int \frac{D\pi}{2\pi} e^{S_E(\beta_0)},
\end{eqnarray}
where, $S_E$ is the Euclidean action in the curved Euclidean space, and 
\begin{multline}
	S_E(\beta_0) =  \int_0^{\beta_0} d\tau \int d^3x  \sqrt{g_e}\\
	\times	\left [ i\pi(\bx,\tau)\frac{1}{\sqrt{h_{00}(\bx)}} D_{0} \phi(\bx,\tau)   -  \sH(\pi(\bx,\tau),\phi(\bx,\tau))  \right ].
\end{multline}
The partition function is then:
\begin{multline}
\label{eq:partitionfunc}
Z(\beta_0) = tr \left [  K(\phi_f,\phi_0,\beta_0) \right ] \\
= tr \left [ \int D\phi \int \frac{D\pi}{2\pi} e^{S_E(\beta_0)} \right ]\\
	 =  \int_{periodic} D\phi \int \frac{D\pi}{2\pi} e^{S_E(\beta_0)}.
\end{multline}
The trace operation implies, $\phi_f = \phi_0 \Rightarrow \phi(\bx,\beta_0) = \phi(\bx,0)$. 
This leads to periodicity being introduced in the curved Euclidean space, just like in the flat Euclidean space. We shall look at the periodicity in more detail in Sec.~\ref{sec:periodicity}.
Equation~\ref{eq:pathintfinal}, plays a pivotal role in establishing the concept of a curved Euclidean space, as an elegant formalism to model spatial thermal gradients. 
For a small temperature variation, i.e., for small values of $s(\bx) - 1$, the partition function, $Z(\beta_0)$, and subsequently, the energy expectation value and entropy, have been calculated for a neutral scalar field in Sec.~\ref{sec:scalarfield}. 

In the current proposal, with the interpretation of the variation of temperature as a variation in the metric, we have,
\begin{equation}
\label{eq:myprop}
\beta(\bx) = \sqrt{h_{00}(\bx)}\beta_0. 
\end{equation}
This is reminiscent of the Ehrenfest and Tolman relation~\cite{ET1,ET2}.   
The Ehrenfest and Tolman relation for a diagonal metric tensor $diag(g_{00}(\bx), 1, 1, 1)$, gives,
\begin{equation}
\label{eq:ET}
\beta(\bx) = \sqrt{-g_{00}(\bx)}\beta_0.
\end{equation}
 Apart from the replacement of $-g_{00}(\bx)$ with $h_{00}(\bx)$, the two relations, given by Eqs.~\ref{eq:myprop} and~\ref{eq:ET}, are similar, mathematically.
The physics, however, is significantly different. Equation~\ref{eq:ET} is interpreted as the appearance of thermal gradients, in a system under global equilibrium, in the presence of a gravitational field. The Lorentzian space-time is curved due to gravity.
On the other hand, Eq.~\ref{eq:myprop} is interpreted as the curvature of the Euclidean space (not Lorentzian) due to thermal gradients. The thermal gradients arise due to heat sources and sinks (Ref. Eqs.~\ref{eq:heateq} and~\ref{eq:zerotempgrad}).
Gravitational field is absent, and the Lorentzian space continues to be flat.
The thermal system is under local thermal equilibrium, and not global equilibrium. 
The similarities and differences lead to physical consequences.
Figures~\ref{fig:l1_20} and ~\ref{fig:u1dirac}, highlight the similarities and differences between the effect of a curved Euclidean and a curved Lorentzian space, on a Dirac fermion eigenstate. This is discussed further in Sec.~\ref{sec:diracgravity}.
\subsection{Generic Hamiltonian}
\label{sec:genham}
Let $H(t)$ be the form of the Hamiltonian, and $\sH(\hat{\pi}(\bx,t),\hat{\phi}(\bx,t),t) \equiv \sH(\bx,t)$, be the corresponding Hamiltonian density.
The analytic continuation, $t \rightarrow i\tau$, gives:
\begin{eqnarray}
	\label{eq:curve}
\nonumber	U(\sH,\beta_0) = \exp\left (-\int\int_0^{s(\bx)\beta_0} \sH(\bx,\tau)d\tau d^3x\right ) \\
	 = \exp \left (-\int \Big \{ F(\bx,s(\bx)\beta_0) - F(\bx,0)\Big \} d^3x \right ) ,
\end{eqnarray}
where $F(\bx,\tau) = \int \sH(\bx,\tau) d\tau$. 
In the case of uniform temperature, the result can be written as:
\begin{eqnarray}
	\label{eq:flat}
\nonumber	U(\sH,\tau_0) = \exp\left (-\int\int_0^{\tau_0} \sH(\bx,\tau)d\tau d^3x\right ) \\
	 = \exp \left (-\int \Big \{ F(\bx,\tau_0) - F(\bx,0)\Big \} d^3x \right ).
\end{eqnarray}
Comparing Eqs.~\ref{eq:curve} and~\ref{eq:flat}, the final result in Eq.~\ref{eq:curve} is obtained by a transformation, $\tau_0 \rightarrow s(\bx)\beta_0$, where $\tau_0$ itself is given by  $i\tau_0 \leftarrow t$. The metric, $ds^2 = -dt^2 + dx^2 + dy^2 + dz^2$, would then transform to:
\begin{equation}
	ds^2 = s(\bx)^2d\beta^2 + dx^2 + dy^2 + dz^2.
\end{equation}
Note: $\bx \equiv (x,y,z)$.
Again, this is a curved space with $h_{00}(\bx) = s(\bx)^2$, indicating a curved Euclideanized space.

The results of Sec.~\ref{sec:timeinvariant} and Sec.~\ref{sec:genham}
indicate, that the evaluation of the partition function in the context of a spatial thermal variation, seamlessly leads to the re-interpretation of the spatial thermal variation as a curved Euclidean space. 

\subsection{Other Thermodynamic Quantities }
With the re-interpretation of the temperature gradient as a system with uniform temperature, $\beta_0$, in a curved Euclidean space, the other thermodynamic quantities can be determined from the partition function $Z(\beta_0)$.
The Helmholtz free energy, $A$, can be defined as:
\begin{equation}
\label{eq:free_energy}
	A(\beta_0) = -\frac{1}{\beta_0}\ln(Z(\beta_0)).
\end{equation}
Since $g_{ij} = \delta_{ij} \forall i,j=1,2,3$, the volume, $\mathbb{V}$, remains the same as a flat Euclidean space. 
The entropy, $S(\beta_0)$, becomes:
\begin{equation}
\label{eq:entropy}
	S(\beta_0) = -\left ( \frac{\partial A}{\partial \Theta_0} \right )_{\mathbb{V}},
\end{equation}
where $\Theta_0 = \frac{1}{\beta_0}$.

\section{Periodicity of the Correlation Function}
\label{sec:periodicity}
For the proposed equivalence between the thermal gradients and the curved Euclidean space to be valid, the periodicity of the correlation function, in a curved Euclidean space, needs to be established. 
We had touched upon the periodicity of the curved Euclidean space in Eq.~\ref{eq:partitionfunc}.
This section explores the periodicity of the correlation function in greater detail.
\subsection{Time invariant Hamiltonian}
\label{sec:corrtimeinvariant}
The case where $\sH(\bx)$ is time invariant, is first considered.
From Eq.~\ref{eq:uhb}, we get,
\begin{eqnarray}
\label{eq:effH}
\nonumber U(\sH,\beta_0)=  \exp\left (-\left [ \int s(\bx)\sH(\bx)d^3x \right ] \beta_0 \right )\\ 
	=  \exp \left (-H' \beta_0 \right ).
\end{eqnarray}
The periodicity of the correlation function is now manifest. An Heisenberg operator, $A$, would evolve as $e^{-H'\beta_0}A(t,i0)e^{H'\beta_0} = A(t,i\beta_0)$. For any two Heisenberg operators, $A$ and $B$, and partition function, $Z (= Z(\beta_0))$, we proceed on lines similar to the standard (KMS) approach~\cite{martin,kubo,Ashok},
\begin{eqnarray}
\nonumber	\langle A_h(t,i0)B_h(t',i0)\rangle 
	= Z^{-1}Tr\left [ e^{-\beta_0 H'}A(t,i0)B(t',i0) \right ] \\
	\nonumber	 = Z^{-1}Tr\left [ e^{-\beta_0 H'}A(t,i0)e^{\beta_0 H'}e^{-\beta_0 H'}B(t',i0) \right ] \\
	\nonumber	 = Z^{-1}Tr\left [ e^{-\beta_0 H'}B(t',i0)A(t,i\beta_0) \right ] \\
	= \langle B(t',i0)A(t,i\beta_0)\rangle~~~~~.
\end{eqnarray}
This result reinforces the interpretation, that, in the curved Euclidean space, the topology is that of a cylinder of uniform radius $\frac{\beta_0}{2\pi}$.
The value of the "constant" $\beta$, in the R.H.S. of Eq.~\ref{eq:qqbeta}, would then be $\beta_0$.
\subsection{Generic Hamiltonian}
\label{sec:corrgeneric}
We now look at a generic Hamiltonian, $H(t) = H_0 + H_I(t)$, where $H_0$ is the free Hamiltonian, and $H_I(t)$ is the interaction part. $H_0$ continues to be time invariant, and the time dependence comes from $H_I(t)$.
Let $\sH(\bx,t)$, $\sH_0(\bx)$ and $\sH_I(\bx,t)$ be the corresponding Hamiltonian densities. 
In the curved Euclidean space, the effective free Hamiltonian  becomes,
 $H_0' = \int s(\bx) \sH_0(\bx) d^3x$. Then, the Interaction picture is given by (as $H_0'$ part of the Hamiltonian is time invariant),
\begin{equation}
	\phi_I(\tau_0) = \exp(\tau_0 H_0')\,\phi(0)\, \exp(-\tau_0 H_0').
\end{equation}
The Heisenberg operator, $\phi_H(\tau_0)$, is given by:
\begin{eqnarray}
\label{eq:heis}
	\nonumber	\phi_H(\tau_0) = \exp \left (\int \int_0^{s(\bx)\tau_0} \sH d\tau d^3x \right )\phi(0)\\
\times \exp\left (-\int \int_0^{s(\bx)\tau_0} \sH d\tau d^3x\right ).
\end{eqnarray}
The relation between the Heisenberg and Interaction picture is then:
\begin{eqnarray}
\label{eq:relint}
	\nonumber	\phi_H(\tau_0) = \exp\left (\int \int_0^{s(\bx)\tau_0} \sH d\tau d^3x\right )\exp(-\tau_0 H_0')\phi_I(\tau_0)\\
\nonumber	 \times \exp\left (\tau_0 H_0'\right )
\nonumber	 \exp\left (-\int \int_0^{s(\bx)\tau_0} \sH d\tau d^3x\right )\\
	 = S^{-1}(\tau_0)\phi_I(\tau_0) S(\tau_0),
\end{eqnarray}
where,
\begin{math}
	S(\tau_0) = \exp\left (\tau_0 H_0'\right ) \exp \left (-\int \int_0^{s(\bx)\tau_0} \sH d\tau d^3x\right ).
\end{math}
For convenience, let us define $R(\tau_0)$ as:
\begin{eqnarray}
\label{eq:R}
\nonumber	R(\tau_0) = \exp\left (-\int \int_0^{s(\bx)\tau_0} \sH(\bx,\tau) d\tau d^3x \right )\\
	= \exp(-\tau_0 H_0')S(\tau_0).
\end{eqnarray}
  The periodicity of the correlation function is now analyzed for $0 < \tau_0 < \beta_0$.
  \begin{multline}
  \label{eq:green_general}
G_{\beta 0}(0,\tau_0) = \langle  T\left [\phi_H(0)\phi_H^{\dagger}(\tau_0)\right ] \rangle \\
= Z^{-1}(\beta_0)Tr\left \{ R(\beta_0) T\left [\phi_H(0)\phi_H^{\dagger}(\tau_0) \right ] \right \}\\
= Z^{-1}(\beta_0)Tr\left \{\pm R(\beta_0)  \phi_H^{\dagger}(\tau_0)\phi_H(0) \right \} \\
= \pm Z^{-1}(\beta_0)Tr\left \{ R(\beta_0)R^{-1}(\beta_0)\phi_H(0) R(\beta_0)  \phi_H^{\dagger}(\tau_0)\right \}.
\end{multline}
It is easily seen from Eq.~\ref{eq:heis}, that,
\begin{eqnarray}
R^{-1}(\beta_0)\phi_H(0) R(\beta_0) 
= \phi_H(\beta_0).
\end{eqnarray}
This gives,
\begin{eqnarray}
	\label{eq:corrpergen}
\nonumber	  G_{\beta 0}(0,\tau_0)= \pm Z^{-1}(\beta_0)Tr\left \{ R(\beta_0)\phi_H(\beta_0) \phi_H^{\dagger}(\tau_0)\right \} \\
	  = \pm G_{\beta 0}(\beta_0,\tau_0),
  \end{eqnarray}
  which manifests the periodicity of the correlation function.
  The $\pm$ sign depends on whether the particle is a fermion or a boson.
  The results of Eqs.~\ref{eq:relint} and~\ref{eq:corrpergen}
  are similar in form, to the results of the uniform temperature case (flat Euclidean space)~\cite{Ashok}.
  The evaluation of the Green's function, however, is a lot more involved as compared to the uniform temperature case.

  The question that now arises is, what happens to periodicity, if we had taken $s(\bx_i) = 1$ instead of $s(0) =1$? Let $s_i(\bx_i) = 1$ and $\beta(\bx) = s_i(\bx_i)\beta_i$. Let's define
\begin{eqnarray}
\label{eq:Rdash}
	R'(\beta_i) = \exp\left (-\int \int_0^{s_i(\bx)\beta_i} \sH(\bx,\tau) d\tau d^3x \right ).
\end{eqnarray}
Equation~\ref{eq:corrpergen} would now read as,
\begin{eqnarray}
	G_{\beta i}(0,\tau_0)= \pm G_{\beta i}(\beta_i,\tau_0).
\end{eqnarray}
This however does not mean that the periodicity has changed from $\beta_0$ to $\beta_i$, by merely redefining $s(x)$.
The function $R'(\beta_i)$ is exactly identical to $R(\beta_0)$ since $s_i(\bx)\beta_i = s(\bx)\beta_0$. 
The correlation function, $G_{\beta i}(0,\tau_0)$, is identical to $G_{\beta 0}(0,\tau_0)$. 
The topological cylinder is non-uniform as shown in Fig.~\ref{fig:single_loop}, and hence possesses a non-uniform periodicity.
Only the reference point related to the non uniform periodicity, has changed from $\beta_0$ to $\beta_i$. 

In summary, it is seen that the periodicity of the correlation function, continues to be maintained in a curved Euclidean space.

\section{Neutral Scalar Field}
\label{sec:scalarfield}
In this section, Eq.~\ref{eq:partitionfunc} is used to determine the partition function for a neutral scalar field, for the case where the temperature variations are small. This requirement translates to $s(\bx)-1~(=\eta(\bx))$ being small. 
Subsequently, the energy, the Helmholtz free energy and entropy are calculated. 

From Eq.~\ref{eq:transamp}, the transition amplitude for a scalar, $\phi(\bx,\tau)$, would be given by:
\begin{multline}
\label{eq:pathintscalar1}
	K(\phi_f,\phi,\beta_0) = 
	\int D\phi \int \frac{D\pi}{2\pi}
	\exp \Bigg \{ \int_0^{\beta_0} d\tau \int d^3x  \sqrt{g_e}\\
	\times	\left [ i\pi(\bx,\tau)\frac{1}{\sqrt{h_{00}(\bx)}}\frac{\partial \phi(\bx,\tau)}{\partial \tau}   -  \sH\left (\pi(\bx,\tau),\phi(\bx,\tau) \right )  \right ]  \Bigg \}.
\end{multline}
The Lagrangian density for a neutral scalar field reads:
\begin{equation}
\label{eq:lagscalar}
	\sL =  \frac{1}{2} \left ( \partial_{\mu}\phi \partial^{\mu}\phi - m^2\phi^2 \right ),
\end{equation}
 and the proper conjugate momenta is, $\pi = \frac{-i}{\sqrt{h_{00}}}\partial_0 \phi$.
Consequently, the Hamiltonian density is:
\begin{equation}
	\sH =   \frac{1}{2} \left ( \pi^2 + (\nabla\phi)^2 + m^2\phi^2 \right ).
\end{equation}
The partition function then reads,
\begin{eqnarray}
\label{eq:pathintscalar2}
\nonumber	Z =		tr K(\phi_f,\phi,\beta_0) = 
	\int_{periodic} D\phi \int \frac{D\pi}{2\pi} \\
\times		\exp \Bigg \{ \int_0^{\beta_0} d\tau \int d^3x  \sqrt{g_e}
\nonumber\Big [ i\pi(\bx,\tau)\frac{1}{\sqrt{h_{00}(\bx)}}\frac{\partial \phi(\bx,\tau)}{\partial \tau}\\
\nonumber	-  \frac{1}{2} \left ( \pi^2 + (\nabla\phi(\bx,\tau))^2 + m^2\phi(\bx,\tau)^2 \right )   \Big ]  \Bigg \}.\\
\end{eqnarray}
To perform the $D\pi$ integrals, we switch to the discretized version and divide the space into $M^3$ cubes of length $\Delta a$ each.
Completing the square and integrating out $\pi(\bx)$, 
we get,
\begin{eqnarray}
\label{eq:pathintscalar3}
	\nonumber		 Z =  \Bigg [ \prod_{N}\prod_{l=1}^{M^3}\left ( \frac{1}{2\pi} \frac{1}{\Delta \tau \Delta a^3\sqrt{g_e(\bx_l)}}  \right )^{1/2} \Bigg ] \\
	\nonumber \times \int_{periodic} D\phi \exp \Bigg \{ \frac{1}{2} \int d\tau \int d^3 x  \sqrt{g_e}\\
	\times	\left [-\frac{1}{h_{00}(\bx)}\frac{\partial \phi}{\partial \tau}\frac{\partial \phi}{\partial \tau}   -   (\nabla\phi)^2 - m^2\phi^2   \right ]  \Bigg \}.
\end{eqnarray}
For the sake of convenience, the continuum domain has been retained for the $\int D\phi$ integrals.
Now, $\phi(\bx,\tau)$ can be decomposed as:
\begin{equation}
\phi(\bx,\tau) =  \frac{1}{\sqrt{\dV \Theta_0}} \sum_i \sum_{n=-\infty}^{\infty}\exp \left \{ i h_{00} p^0 \tau \right \} e^{i\bp_i.\bx}\phi_{n,\bp},\\
\end{equation}
with, $p^0 = h^{00}C_n$, being the discrete Matsubara frequency, and  
$C_n = \frac{2\pi n}{\beta_0}$ for integer $n$.
Also, $\Theta_0 = \frac{1}{\beta_0}$, is the "constant" temperature in the curved Euclidean space, and $\dV$ is the volume.
This, finally, provides the Fourier decomposition as:
\begin{equation}
	\label{eq:phiexp}
\phi(\bx,\tau) = \sqrt{\frac{\beta_0}{\dV}} \sum_i\sum_{n=-\infty}^{\infty}e^{iC_n\tau} e^{i\bp_i.\bx}\phi_{n,\bp_i}.
\end{equation}
The Matsubara frequency has been discretized due to the periodicity condition imposed on $\phi$.
The above form of the Matsubara frequency, keeps the phase factor, $\exp(iC_n\tau)$, independent of $\bx$, 
and in Sec.~\ref{sec:dirac}, can be seen to also solve the Dirac equation for a spinor. 
A more detailed explanation of the Matsubara frequency, in the context of a curved Euclidean space, can be found in Sec.~\ref{sec:matsubara}.

Substituting $\phi(\bx,\tau)$ from Eq.~\ref{eq:phiexp}, the exponent in Eq.~\ref{eq:pathintscalar3} can be written as
\begin{multline}
\label{eq:pathintscalar4}
	\int d\tau \int d^3x \sqrt{g_{e}} \left [-\frac{1}{h_{00}(\bx)}\frac{\partial \phi}{\partial \tau}\frac{\partial \phi}{\partial \tau}   -   (\nabla\phi)^2 - m^2\phi^2   \right ] \\ 
= -\sum_{n=-\infty}^{\infty} \frac{\beta_0^2}{\dV}\int d^3x \sum_i\sum_j 
\Big ( \frac{C_n^2}{s(\bx)} + s(\bx)\bp_i \bp_j\\
	+ s(\bx) m^2  \Big ) 
e^{i(\bp_i - \bp_j).\bx} \phi_{n,\bp_j}^* \phi_{n,\bp_i}.
\end{multline}
Let $s(\bx) = 1 + \eta(\bx)$. If $\eta(\bx)$ is small, then $\frac{1}{s(\bx)} \approx 1 - \eta(\bx)$. 
Expand $\eta(\bx)$ in terms of its Fourier components, $\eta(\bx) = \frac{1}{\dV} \sum_q  \eta(\bq) \exp(i\bq.\bx)$. After carrying out the $\int d^3x$ integral, and summing over $\bq$ to eliminate the resulting $\delta$ function, one obtains, 
\begin{multline}
\label{eq:pathintscalar5}
\int d\tau \int d^3x \sqrt{g_{e}} \left [-\frac{1}{h_{00}(\bx)}\frac{\partial \phi}{\partial \tau}\frac{\partial \phi}{\partial \tau}   -   (\nabla\phi)^2 - m^2\phi^2   \right ] \\ 
	= -\sum_{n=-\infty}^{\infty} \beta_0^2 \sum_i \sum_j
\phi_{n,\bp_j}^* \Bigg [ (C_n^2 + \bp_i \bp_j +  m^2)\delta_{ij}\\
	+  \frac{1}{\dV}\eta_f(\bp_j - \bp_i) (-C_n^2 +\bp_i \bp_j +m^2)  \Bigg ]  \phi_{n,\bp_i},
\end{multline}
where, the $\delta$ function, in the discretized space, is defined as:
\begin{equation}
	\nonumber	\delta_{ij} = 
	\begin{cases}
		1 & if~ \bp_i = \bp_j,\\
		0 & \it otherwise.
	\end{cases}
\end{equation}

Let $J$ be the Jacobian for converting the integral measure, $\int D\phi(\bx,\tau)$, to the measure $\int D \phi_{n,\bp}$.
The Jacobian, $J$, is lumped along with the factor $\prod_N\prod_l\left ( \frac{1}{\Delta \tau \Delta a^3}  \right )^{1/2}$, into an overall factor, $\sN$. The factor, $\sN$, does not affect the dynamics of the system~\cite{kapusta, Nbeta}. The new factor, $\frac{1}{\sqrt{g_e(\bx_l)}}$, is independent of $\beta_0$ and $\dV$. Consequently, it is irrelevant to the thermodynamics, and is clubbed into $\sN$.

The volume, $\dV$, can be represented in the discrete momentum domain as, $\dV = \left ( \frac{2\pi}{\Delta p} \right )^3$.
Substituting the expression in Eq.~\ref{eq:pathintscalar5}, inside the exponent of Eq.~\ref{eq:pathintscalar3}, we get,
\begin{eqnarray}
\label{eq:pathintscalar7}
	\nonumber		   Z =   \sN \left ( \frac{1}{2\pi}   \right )^{\frac{NM^3}{2}}   \int D\phi_{n,\bp} \exp \Bigg \{ \frac{-1}{2} \sum_{n=-\infty}^{\infty} \sum_i \sum_j \\
	\nonumber \times	\phi_{n,\bp_j}^* \Bigg [\beta_0^2 (C_n^2 + \bp_i \bp_j +  m^2)\delta_{ij} \\
	\nonumber\times	\Big \{ I_{ij} + \frac{\delta_{ij}}{C_n^2 + \bp_i\bp_j +m^2}\eta_f(\bp_j - \bp_i)\\
\times	\frac{\Delta p^3}{(2\pi)^3}(-C_n^2 +\bp_i \bp_j +m^2) \Big \} 
	\Bigg ]  \phi_{n,\bp_i} \Bigg \}.
\end{eqnarray}
Note that the matrix, $\frac{\delta_{ij}}{C_n^2 + \bp_i\bp_j +m^2}$, is the inverse of the diagonal matrix, $(C_n^2 + \bp_i \bp_j +  m^2)\delta_{ij}$, and there is no implied sum over repeated indices.
Finally, evaluating the $\int D\phi_{\bp}$ Gaussian integrals,
\begin{eqnarray}
\label{eq:pathintscalar8}
\nonumber		   Z = \sN  \prod_{n=-\infty}^{\infty}  \Bigg \{ 
	\det \Bigg [ \beta_0^2 (C_n^2 + \bp_i \bp_j +  m^2)\delta_{ij} \\
	\nonumber \times	\Big \{ I_{ij} + \frac{\delta_{ij}}{C_n^2 + \bp_i\bp_j +m^2}\eta_f(\bp_j - \bp_i)\\
\times	\frac{\Delta p^3}{(2\pi)^3}(-C_n^2 +\bp_i \bp_j +m^2) \Big \} 
	\Bigg ]  \Bigg \}^{-1/2}.
\end{eqnarray}

To calculate $\ln(Z)$, 
we make use of the identities, $\det(M) = \exp(tr \ln(M))$, and $\ln(1+X) = \sum_r\frac{-1^{r-1}}{r}X^r$. After dropping the irrelevant factor, $\sN$, we get,

\begin{multline}
\label{eq:pathintscalar10}
	\ln(Z) =  \sum_{n=-\infty}^{\infty}  \sum_{i} \Big [ 
	- \frac{1}{2} \ln \left ( \beta_0^2(C_n^2 + \bp_i^2 +  m^2) \right ) \Big ] \\
	-\sum_{n=-\infty}^{\infty}\frac{1}{2} tr \sum_r \frac{-1^{r-1}}{r} \Big \{ \frac{\delta_{ij}}{C_n^2 + \bp_i\bp_j +m^2}\eta_f(\bp_j - \bp_i)\\
\times	\frac{\Delta p^3}{(2\pi)^3}(-C_n^2 +\bp_i \bp_j +m^2) 
	\Big \}^{r}.
\end{multline}

The trace operation is on the indices, $i,j$.
We substitute $C_n = \frac{2\pi n}{\beta_0}$, 
Approximating the power series in $r$ to the first order, and  differentiating w.r.t. $\beta_0$,
and finally, going over to the continuum limit in the momentum space, 
\begin{eqnarray}
\label{eq:pathintscalar12}
	\nonumber		  -\frac{\partial \ln(Z)}{\partial \beta_0} = 
		\frac{1}{2}   \frac{(2\pi)^3}{(\Delta p)^3} \int \frac{d^3p}{(2\pi)^3} \sum_n \Big [ \frac{ \frac{\beta_0}{2\pi^2}\omega_p^2}{(n^2 + \left (\frac{\beta_0}{2\pi} \right )^2 \omega_p^2 )} \Big ] \\
	+\frac{1}{2} \eta_f(0) \int \frac{d^3p}{(2\pi)^3}
	 \Bigg [
		 \nonumber		 \frac{\partial}{\partial \beta_0}	 \sum_n \Big \{ \frac{2 \left ( \frac{\beta_0}{2\pi} \right )^2 \omega_p^2} {n^2 + \left (\frac{\beta_0}{2\pi} \right )^2 \omega_p^2 } -1\Big \} 
	  \Bigg ],\\
\end{eqnarray}
where, $\omega_p = \sqrt{\bp^2 + m^2}$. The above is easily evaluated to obtain,
\begin{eqnarray}
\label{eq:pathintscalar14}
	\nonumber -\frac{\partial \ln(Z)}{\partial \beta_0} = 
		 \int \frac{d^3p}{(2\pi)^3} \Bigg [ \dV \frac{\omega_p}{2} \coth\left (\frac{\beta_0\omega_p}{2} \right ) \\
	+ \eta_f(0) \left \{ \frac{\partial}{\partial \beta_0}\left ( \frac{\beta_0 \omega_p}{2} \coth\left( \frac{\beta_0 \omega_p}{2} \right ) \right ) \right \}\Bigg ].
\end{eqnarray}
The factor, $\eta_f(0)  = \left (\frac{\dV}{\int d^3x} \right )\int \eta(\bx) d^3x = \dV \langle \eta(\bx) \rangle$, where $\langle \eta(\bx) \rangle$ is the mean value of $\eta(\bx)$.
Finally, after carrying out the explicit differentiation in Eq.~\ref{eq:pathintscalar14}, we obtain the expected value of energy, $\langle E \rangle$.
\begin{multline}
\label{eq:pathintscalar15}
	\langle E \rangle =	  -\frac{\partial \ln(Z)}{\partial \beta_0} = 
		 \dV \int \frac{d^3p}{(2\pi)^3} \Bigg [ \frac{\omega_p}{2} \coth\left (\frac{\beta_0\omega_p}{2} \right ) \\
+ \langle \eta(\bx) \rangle 
	\Big \{	  \frac{\omega_p}{2} \coth\left (\frac{\beta_0\omega_p}{2} \right ) + \frac{\beta_0 \omega_p^2}{4} 
	 \cosech^2 \left ( \frac{\beta_0\omega_p}{2} \right )  \Big \} \Bigg ].
\end{multline}
It is seen that the contribution to the energy, $\langle E\rangle$, due to the variation in the temperature scale, $\eta(\bx)$, is proportional to $\langle \eta(\bx)\rangle$, to a first order approximation. This corroborates with a naive expectation.

Now, $\ln(Z)$ can easily be evaluated by integrating Eq.~\ref{eq:pathintscalar14} w.r.t. $\beta_0$, and discarding the ($\beta_0$ independent) constant of integration.
\begin{multline}
\label{eq:pathintscalar16}
	\ln(Z) = 
		-\dV \int \frac{d^3p}{(2\pi)^3} \Bigg [ \frac{\beta_0 \omega_p}{2}  + \ln\left \{1 - \exp \left (-\beta_0\omega_p \right ) \right \} \\
	+ \langle \eta(\bx) \rangle \left ( \frac{\beta_0 \omega_p}{2} \right ) \coth\left( \frac{\beta_0 \omega_p}{2} \right )  \Bigg ].
\end{multline}
The Helmholtz free energy, $A(\beta_0)$, is obtained by using the relation specified in Eq.~\ref{eq:free_energy}:
\begin{multline}
\label{eq:pathintscalar17}
	A(\beta_0) = -\frac{1}{\beta_0}\ln(Z) =  \\
\dV \int \frac{d^3p}{(2\pi)^3} \Bigg [ \frac{\omega_p}{2}  + \frac{1}{\beta_0}\ln\left \{1 - \exp \left (-\beta_0\omega_p \right ) \right \} \\
	+ \langle \eta(\bx) \rangle  \frac{\omega_p}{2} \coth\left( \frac{\beta_0 \omega_p}{2} \right ) \Bigg ].
\end{multline}
By using the relation in Eq.~\ref{eq:entropy},
the entropy, $S(\beta_0)$, can be obtained from Eq.~\ref{eq:pathintscalar17}, 
\begin{equation}
\label{eq:scalarentropy}
	\nonumber	S(\beta_0) = -\frac{\partial}{\partial \Theta_0} \left ( A(\beta_0) \right ) = \beta_0^2 \frac{\partial A(\beta_0) }{\partial \beta_0}.
\end{equation}
This gives,
\begin{multline}
	S(\beta_0) = \dV \int \frac{d^3p}{(2\pi)^3} \Bigg [ -\ln\left \{1 - \exp \left (-\beta_0\omega_p \right ) \right \}\\
+ \beta_0 \omega_p \frac{\exp(-\beta_0\omega_p)}{1-\exp(-\beta_0\omega_p)}\\
	+ \langle \eta(\bx) \rangle  \left ( \frac{\beta_0\omega_p}{2} \right )^2 \cosech^2\left( \frac{\beta_0 \omega_p}{2} \right )  \Bigg ].
\end{multline}
The Ricci scalar curvature, $R$, has not been included in the Lagrangian in Eq.~\ref{eq:lagscalar}. But the above analysis can be extended in a straightforward manner, albeit involving more tedious calculations, to include $R$.
For the Lagrangian,
\begin{equation}
\label{eq:lagscalarR}
        \sL =  \frac{1}{2} \left ( \partial_{\mu}\phi \partial^{\mu}\phi - m^2\phi^2 - \zeta R \right ),
\end{equation}
Eq.~\ref{eq:pathintscalar15} becomes,
\begin{multline}
\label{eq:pathintscalarR1}
        \langle E \rangle =       -\frac{\partial \ln(Z)}{\partial \beta_0} =
                 \dV \int \frac{d^3p}{(2\pi)^3} \Bigg [ \frac{\omega_p}{2} \coth\left (\frac{\beta_0\omega_p}{2} \right ) \\
	+ \left ( \langle \eta(\bx) \rangle  + \frac{\zeta}{2\omega_p^2}\langle s(\bx) R(\bx)\rangle \right ) \\
\times        \Big \{   \frac{\omega_p}{2} \coth\left (\frac{\beta_0\omega_p}{2} \right ) + \frac{\beta_0 \omega_p^2}{4}
         \cosech^2 \left ( \frac{\beta_0\omega_p}{2} \right )  \Big \} \Bigg ],
\end{multline}
where, $\langle s(\bx)R(\bx)\rangle = \frac{\int d^3x s(\bx)R(\bx)}{\int d^3x}$.
\section{Classical system and the Geodesic Equation}
\label{sec:geodesic}
This section explores the applicability of the equivalence between spatial thermal gradients, and curved Euclidean space, in a classical scenario.
We determine the equations describing the kinetics of a massive test particle in thermal medium with spatial thermal gradients, and compare it with the geodesic equations of the curved Euclidean space.
Besides, this provides an interpretation of the geodesics in a curved Euclidean space.
\subsection{The Geodesic Equation}
Let us consider a candidate metric, where the temperature varies across the two dimensional $x-z$ plane.
\begin{equation}
\label{eq:metric2D}
	ds^2 = h_{00}(x,z)dt^2 - dx^2 - dy^2 - dz^2,
\end{equation}
with $t = i\beta$.
The temperature variation is given by $\Theta(x,z) = \frac{1}{\beta_0\sqrt{h_{00}(x,z)}}$.

Two of the geodesic equations for the above metric include,
\begin{eqnarray}
	\label{eq:geodesic}
	\nonumber	\left (\frac{\partial \beta}{\partial \lambda} \right )^2 \frac{\partial h_{00}(x,z)}{\partial x}  - 2\frac{\partial^2 x}{\partial \lambda^2} = 0,\\
	2\frac{\partial^2 z}{\partial \lambda^2} - \left ( \frac{\partial \beta}{\partial \lambda} \right )^2 \frac{\partial h_{00}(x,z)}{\partial z} = 0,
\end{eqnarray}
where, $\lambda$ is an affine parameter.
Eliminating $\left (\frac{\partial \beta}{\partial \lambda} \right )^2$ 
between the set of equations in Eq.~\ref{eq:geodesic}, we obtain,
\begin{equation}
	\label{eq:xzcurve}
	\frac{\partial h_{00}}{\partial x} \frac{\partial^2 z}{\partial \lambda^2} = \frac{\partial h_{00}}{\partial z} \frac{\partial^2 x}{\partial \lambda^2}.
\end{equation}
Equation~\ref{eq:xzcurve} indicates that, for any non-zero $\frac{\partial h_{00}}{\partial x}$ and $\frac{\partial h_{00}}{\partial z}$, the particle will bend, and the geodesic is not straight.
For photons, the above may be seen equivalent to bending of light in a medium of varying refractive index. 
Since no quantum condition has been imposed in arriving at the geodesic equation, it is expected to be valid in the classical limit, where the deBroglie wavelength of the particle is much smaller than the space-time curvature.
This section makes use of the geodesic equation to provide a physical validation of the theory, albeit, at a classical level.
\subsection{Interpretation at a classical level}
We now look at the possible origins of the geodesic curve, given by Eq.~\ref{eq:xzcurve}, from a kinetic theory point of view, for a massive particle.
We study the propagation of a massive test particle in a plasma.
Let $\langle v^2(x,z)\rangle$ be the expected r.m.s. velocity of a constituent particle of the medium at a spatial location, $(x,z)$.
Classically,
\begin{math}
	\frac{1}{2}m_p\langle v^2 \rangle = \frac{n_p}{2}k\Theta,
\end{math}
where $n_p$ and $m_p$ are the degrees of freedom and the effective mass of the constituent plasma particle, and $\Theta$ is the temperature.
This means,
\begin{math}
	\sqrt{\langle v^2(x,z) \rangle } = \sqrt{\frac{n_p k}{m_p}} \sqrt{\Theta} = k_0\sqrt{\Theta}.
\end{math}
Rewriting $\Theta$ in terms of the metric $h_{00}(x,z)$,
\begin{equation}
	\sqrt{\langle v^2(x,z) \rangle } = \frac{k'}{h_{00}(x,z)^{1/4}},
\end{equation}
with $k' = \frac{k_0}{\sqrt{\beta_0}}$.
The gradient in the velocity along the $x$-direction, would then become,
\begin{equation}
	\partial_x\sqrt{\langle v^2(x,z) \rangle } = \frac{k''}{h_{00}(x,z)^{5/4}}h_{00x},
\end{equation}
where, $h_{00x} = \frac{\partial h_{00}(x,z)}{\partial x}$.
The variation in r.m.s. velocity of the plasma constituents will bring about a change in momentum of the test particle in the plasma.  
The momentum change of the test particle over a unit distance $\Delta x = \Delta l$, is given by: 
\begin{equation}
	\Delta p_x \propto \partial_x\sqrt{\langle v^2(x,z) \rangle } \Delta l.
\end{equation}
Acceleration of the test particle would be proportional to the rate of change of momentum. This gives, $a_x  \propto  \frac{\Delta p_x}{\Delta t}$. Under conditions of local thermodynamic equilibrium and stationarity, $\partial_x\sqrt{\langle v^2(x,z) \rangle }$ is static in time.
Furthermore, the time variable, $t$, is analytically continued and replaced by the affine parameter $\lambda$ (i.e. $\frac{\partial}{\partial t} \propto i\frac{\partial}{\partial \lambda}$).  This finally gives,
\begin{equation}
	\label{eq:accx}
	a_x  \propto i\partial_x\sqrt{\langle v^2(x,z) \rangle } \frac{\Delta l}{\Delta \lambda} = \frac{ik'' h_{00x}}{h_{00}(x,z)^{5/4}} \frac{\Delta l}{\Delta \lambda}.
\end{equation}
In the $z$-direction, we again calculate the momentum change, $\Delta p_z$, over the same (in magnitude) unit distance, $\Delta z = \Delta l$. This gives,
\begin{equation}
	\label{eq:accz}
	a_z  \propto  \frac{ik'' h_{00z}}{h_{00}(x,z)^{5/4}} \frac{\Delta l}{\Delta \lambda}.
\end{equation}
Rewriting, $a_x = \frac{\partial^2x}{\partial \lambda^2}$ and $a_z = \frac{\partial^2z}{\partial \lambda^2}$, and taking the ratio of Eqs.~\ref{eq:accx} and~\ref{eq:accz}, we get,
\begin{equation}
	\frac{1}{h_{00x}}\frac{\partial^2 x}{\partial \lambda^2} = \frac{1}{h_{00z}}\frac{\partial^2 z}{\partial \lambda^2},
\end{equation}
which is nothing but the geodesic equation, Eq.~\ref{eq:xzcurve}.

   What has been achieved here, is that the curved path taken by the test particle, determined by the kinetic considerations after an analytic continuation ($\frac{\partial }{\partial t} \rightarrow i\frac{\partial}{\partial \lambda}$), is seen to be the geodesic in the curved Euclidean space.
This result plays an important role, and is a step towards establishing the physical validity of the theory.

A note: Equation~\ref{eq:metric2D} begins with the signature, (+,-,-,-), and after the analytical continuation, $t \rightarrow i\beta$, it becomes (-,-,-,-). 
One could begin with (-,+,+,+), leading to a more pleasing, (+,+,+,+), after the analytic continuation. The geodesic (Eq.~\ref{eq:geodesic}) remains the same in either case.
The former signature (+,-,-,-) has been chosen to be consistent with the convention used in particle physics, in order to facilitate solving the Dirac equation in Sec.~\ref{sec:dirac}.

\section{Dirac Spinor Eigenstate Transfiguration}
\label{sec:dirac}
Having justified the concept of a curved Euclidean space in a generic field theoretic and a classical setup, in the earlier sections, we now attempt to understand the important properties of the solutions of the relativistic Dirac equation.
This section models an external Dirac spinor, with energy, $E$, and mass, $m_g$, propagating through a thermal medium with spatial thermal gradients. 
The external Dirac spinor is modeled via the Dirac equation.
The above setup helps to elucidate the basic quirk(s) of the theory, and provides a mechanism for the physical validation of the mathematical framework of the curved Euclidean space. 
These properties should carry over to the field theoretic aspects as well.
To elaborate, let the solutions of the Dirac equation, $\psi_n(\beta,z)$, constitute the complete set of eigenfunctions (or eigenstates). Consequently, $\psi_n(\beta,z)$ forms a basis set, and any solution can then be written as $\psi(\beta,z) = \sum_n a_n\psi_n(\beta,z)$, with $a_n$ being arbitrary constants. For a field theoretic approach, we promote $a_n$ to an operator, satisfying the standard anti-commutation relations. Since $\psi_n(\beta,z)$ forms the complete solution set, we can express the field theoretic solution, $\Psi(\beta,z)$, as $\Psi(\beta,z) = \sum_n  a_n \psi_n(\beta,z)$, with $a^{\dagger}_n$ and $a_n$, now becoming the creation and annihilation operators. 
As a side note, the antiparticle wave-functions have been omitted for both $\psi(\beta,x)$ and $\Psi(\beta,x)$, which is but a matter of detail.
Any fundamental property or quirks, that $\psi_n(z)$ exhibits, would be carried over to the field theoretic solutions as well. Therefore, investigating the properties of the basis function, $\psi_n(z)$, in a given curved space, is fundamental to understanding the physics of the quantum theory in that curved space.  
\subsection{Formulation}
\label{sec:diracform}
A candidate Euclidean metric representing spatial temperature variation along the 1 dimensional z-direction is:
\begin{equation}
\label{eq:metric}
	ds^2 = h_{00}(z)dt^2 - dx^2 - dy^2 - dz^2,
\end{equation}
with $h_{00}(z) = V(z)$, and $t = i\tau$.

The Ricci scalar curvature for the metric in Eq.~\ref{eq:metric} is given by,
\begin{math}
	R = \frac{2VV'' - V'^2}{2V^2},
\end{math}
where  the $'$ refers to differentiation w.r.t $z$.

The Dirac equation in curved space time is given by~\cite{dirac0,dirac,dirac_rus,book2}:
\begin{equation}
\label{eq:dirac}
	i\gamma^a e^{\mu}_a D_{\mu}\psi - m\psi = 0,
\end{equation}
where,
\begin{itemize}
\item \begin{math}
	D_{\mu} = \partial_{\mu} - \frac{i}{4}\omega_{\mu}^{ab}\sigma_{ab},
\end{math}, with $\omega_{\mu}^{ab}$ = spin connection,
\item $\sigma_{ab} = \frac{i}{2}[\gamma_a,\gamma_b]$, and $\gamma_a$ being the Dirac matrices in flat space-time, and 
\item $e_{\mu}^a$ is the vierbein.
\item In vacuum (0 temperature), $m$ is the gravitational mass. But in the proposed thermal system, an effective mass, $m$, has been used. 
\end{itemize}
	In the notation used, $\mu$, $\nu$, etc., are indices in curved space-time, and $a$, $b$, etc., are indices in flat space-time.

	The mass, $m$, is not the gravitational mass, $m_g$, but an effective mass, $m = m_g - i\gamma^0 E$, where $E$ is the energy  eigenvalue of a stationary state. The effective mass has been so chosen, so that this equation boils down to the Dirac Equation in vacuum, whose solution is the stationary plane wave, $exp\{-i(Et - \vec{p}.\vec{x})\}$, when the temperature $\Theta (\equiv \frac{1}{\tau}) \rightarrow 0$.
	An alternate viewpoint could be that, $\gamma^0E\psi$ is the result of the operation, $\gamma^0\partial_t \psi$, for a time invariant system. This is revisited in Sec.~\ref{sec:interp}.

		We now solve the Dirac equation given by Eq.~\ref{eq:dirac}. 
		The coordinate axes used are $(i\tau,x,y,z)$. The term, $\frac{-i}{4}\omega_{\mu}^{ab}\sigma_{ab}$, for the metric in Eq.~\ref{eq:metric}, evaluates to $\frac{f(z)}{8}\gamma^0\gamma^3$, with $\gamma^{0}$ and $\gamma^3$ being the Dirac matrices in flat space, and $f(z) = \frac{-2V'(z)}{\sqrt{V(z)}}$.
We take the solution to be of the form
\begin{equation}
	\label{eq:massive_psi}
	\psi = u(z) \exp(-ih_{00}p^0\tau), 
\end{equation}
		where the spinor $u(z)$ is given by:
		\begin{equation}
			\label{eq:spinor_u}
			u(z) = \left( \begin{array}{c} (\zeta_1(z) + i\eta_1(z)) \zeta_0 \\ (\zeta_2(z) + i\eta_2(z)) \zeta_0 \end{array} \right).
		\end{equation}
		Here, $\zeta_0$ = 
		$\left( \begin{array}{c}
			1\\
			0\\
		\end{array} \right )$ or 
		$\left (\begin{array}{c}
			0\\
			1\\
		\end{array} \right )$, i.e., $\zeta_0$ indicates a spin up or spin down particle. 
Substituting in Eq.~\ref{eq:dirac}, and equating both the real and imaginary parts to 0, we get:
\begin{eqnarray}
\label{eq:partialsl}
\nonumber	-k_2 u_1 -  s u_1'  + im_g u_2 = 0,\\
		-k_1 u_2 +  s u_2'  + im_g u_1 = 0,
\end{eqnarray}

where, 
\begin{itemize}
	\item $u_j = \zeta_j + i\eta_j$; j = 1,2,
	\item $s$ is the eigenvalue of the Pauli $\sigma^3$ matrix,
	\item $k_1 = iE + \frac{-C_n - f(z)s/8}{\sqrt{V(z)}}$, and $k_2 = iE + \frac{-C_n + f(z)s/8}{\sqrt{V(z)}}$.
\end{itemize}
It is to be noted that $m_g$ is the gravitational mass.
In coming up with the equations in Eq.~\ref{eq:partialsl}, we have taken $p^0 = \frac{C_n}{V(z)}$, where, $C_n = \frac{(2n+1)\pi}{\beta_0}$, n = 0,1,2,...; $\beta(z)$ is the inverse temperature and $\beta_0 = \beta(0)$.
In flat space-time, with $V(z) = V_0=$ constant, $\frac{C_n}{V_0}$ would be the Matsubara frequency. 
In this article, $C_n$ itself is sometimes referred to as the Matsubara frequency. 
This choice of $p^0$, keeps the exponent in Eq.~\ref{eq:massive_psi} independent of $z$, and is also seen to solve the Dirac equation.
The anti-periodicity of $\psi$ is discussed in Sec.~\ref{sec:matsubara}.

A simple algebraic manipulation of the equations in Eq.~\ref{eq:partialsl}, results in: 
\begin{eqnarray}
\label{eq:secondorder}
	\nonumber	u_1'' + \frac{f}{4\sqrt{V}}u_1' + (-k_1 k_2 + sk_2' - m_g^2)u_1 = 0,\\
	u_2'' + \frac{f}{4\sqrt{V}}u_2' + (-k_1 k_2 - sk_1' - m_g^2)u_2 = 0.
\end{eqnarray}
  \subsection{Matsubara frequency}
  \label{sec:matsubara}
	We now explore the anti-periodicity of the fermionic wave-function, $\psi$, around the topological cylinder at all points in space.
	As discussed earlier, let $C_n = \frac{(2n+1)\pi}{\beta(0)}$.
Then, $\exp\left \{-ih_{00}(z)p^0(z)\tau  \right \}  = \exp\left \{-i(2n+1)\pi \right \} $  at $\tau = \beta(0)$, which is independent of $z$. In other words, the anti-periodicity of the fermion wave-function is retained at all values of $z$. 
	This also determines that, $\beta = \beta(0)$, would be the value used in R.H.S. of Eq.~\ref{eq:qqbeta}. 
	This result corroborates with the result obtained in Sec.~\ref{sec:periodicity}.
	We note that the inverse temperature at a point, $z$, is given by "proper" $\beta$ = $\beta_p = \sqrt{h_{00}(z)}\tau$. A "proper" Matsubara frequency could then be given as, $\omega_p = \sqrt{h_{00}}p^0 = \frac{C_n}{\sqrt{V(z)}}$.

\subsection{Thermal Bath with zero temperature gradient}
To gain an insight into Eq.~\ref{eq:secondorder}, we first solve Eq.~\ref{eq:secondorder} in flat space, i.e., when the temperature gradient is 0. 
This would provide the set of baseline solution, $u_1^{flat}$ and $u_2^{flat}$. 
When, the gradient $V'(z)=0$, and $V(z) = V_0$, the equations in Eq.~\ref{eq:secondorder} simplify to
\begin{eqnarray}
	\label{eq:nograd}
	\nonumber	u_1'' +  ( -(iE -\frac{C_n}{\sqrt{V_0}})^2  - m_g^2)u_1 = 0,\\
		u_2'' +  ( -(iE -\frac{C_n}{\sqrt{V_0}})^2  - m_g^2)u_2 = 0,
\end{eqnarray}
	whose solution is given by: 
	\begin{eqnarray}
		\label{eq:flatsol}
		\nonumber		u_1^{flat} = A_1\exp\Bigg ( \frac{1}{\sqrt{2}} \Big \{\sqrt{-Q_r + \sqrt{Q_r^2 + Q_i^2}  }\\
\nonumber		+i\sqrt{Q_r + \sqrt{Q_r^2 + Q_i^2} } \Big \}z  \Bigg )\\
\nonumber		+	 A_2\exp\Bigg ( \frac{-1}{\sqrt{2}} \Big \{ \sqrt{-Q_r + \sqrt{Q_r^2 + Q_i^2}  }\\
		+i\sqrt{Q_r + \sqrt{Q_r^2 + Q_i^2} } \Big \}z \Bigg ),
	\end{eqnarray}
where, $Q_r =  ( E^2 - m_g^2 ) - \frac{C_n^2}{V_0}  $ and  $Q_i = \frac{2EC_n}{\sqrt{V_0}}$.
	As the temperature  $\Theta \rightarrow 0$, $V_0 \rightarrow \infty$, the resulting solution $ e^{\left  (i\sqrt{Q_r}z \right )} \rightarrow e^{\left \{i \left (\sqrt{E^2 -  m_g^2}\right )z \right \}}$. This is nothing but the spatial part of the plane wave $e^{-i(Et - \vec{p}.\vec{x})}$, namely, $e^{i\vec{p}.\vec{x}}$, where $\vec{p}$, is the 3-momentum and $p^2= (E^2 - m_g^2)$.
	For finite temperatures, one can infer the following two observations
	\begin{itemize}
		\item The resultant 3-momentum in the thermal bath, $q_i$, is given by:
\begin{equation} 
\nonumber q_i = \sqrt{\frac{1}{2}(Q_r + \sqrt{Q_r^2 + Q_i^2})}, 
\end{equation} 
which is less than the 3-momentum in vacuum, namely, $\sqrt{  (E^2 - m_g^2)}$.
		\item There is a decay factor, $q_r$, given by:
\begin{equation}
\nonumber q_r = \sqrt{\frac{1}{2}(-Q_r + \sqrt{Q_r^2 + Q_i^2})}.
\end{equation}
	\end{itemize}
The fermion then acts as a particle, with energy $E$, 3-momentum $q_i$, and mass $\sqrt{E^2 - q_i^2}$.


It may be possible then to reduce  Eq.~\ref{eq:secondorder} to the form of Eq.~\ref{eq:nograd}, i.e., without the first derivative term, $\frac{fs}{4\sqrt{V(z)}}$. 
In the form present in Eq.~\ref{eq:nograd}, the second derivative can be used to provide the square magnitude of the 3-momentum. 
This would be the subject matter of discussion in the Sec.~\ref{sec:tempgrad}.
As a reminder, we have seen that, in the case of nil temperature gradient, the $\frac{fs}{4\sqrt{V(z)}}$ term disappears, and $q_i = Imag(\sqrt{\frac{u_1^{''}}{u_1}})$ (=  $Imag(\sqrt{\frac{u_2^{''}}{u_2}})$) directly provides the 3-momentum observable. 

\subsection{Thermal Bath with a temperature gradient}
\label{sec:tempgrad}
We now look at the scenario, where there is a temperature gradient.

Define $h(z)$ and $l(z)$ as follows:
\begin{eqnarray}
	\label{eq:hz}
\nonumber	h(z) =  \exp \left (-\frac{1}{2}\int \frac{f(z)}{4\sqrt{V(z)}} dz \right )\\
=  \exp \left (-\frac{1}{4}\int \frac{V'(z)}{V(z)} dz \right )
	 =  C' V(z)^{-1/4}, 
\end{eqnarray}
where $C'$ is a constant of integration and,
\begin{equation}
	\label{eq:li}
	l_i(z) = h(z)u_i(z), i=1,2.
\end{equation}
The function, $h(z)$, depends on the ratio $\frac{V'(z)}{V(z)}$, and hence is scale independent. In the current context, scale implies the temperature scale, $\beta_0$. It is a factor which depends purely on the thermal gradient.
We express $h(z)$ as:
to highlight the fact that $d(z)$ is scale independent.
Equation~\ref{eq:li} indicates that the function $h(z)$ becomes then an exponential decay or recovery factor for $u_i(z)$, and depends only on the curvature.
The set of equations \ref{eq:secondorder}, can then be cast in the form,
\begin{eqnarray}
\label{eq:leq}
\nonumber	\frac{l_1^{''}}{l_1} = - {(-k_1k_2 + sk_2^{'} - m^2) - h^{''}(z))},\\
	\frac{l_2^{''}}{l_2} = - {(-k_1k_2 - sk_1^{'} - m^2) - h^{''}(z))}.
\end{eqnarray}
It can be observed that this is of a similar form as Eq.~\ref{eq:nograd}.
The exact solution to the above has to be determined numerically. However, if the temperature gradient is very small, one can further analyze Eq.~\ref{eq:leq} analytically.
\subsubsection{Small temperature gradient}
\label{sec:smallgrad}
The temperature gradient being small implies that 
	the derivative, $\frac{\partial}{\partial z} \left \{-{(-k_1k_2 + sk_2^{'} - m^2) - h^{''}(z)} \right \} \sim 0$.
It can be seen that,
\begin{itemize}
	\item  $\frac{l_1^{''}}{l_1}$ = $\left \{-{(-k_1k_    2 + sk_2^{'} - m_g^2) - h^{''}(z)} \right \} = \left \{-(E^2 - m_g^2) + \frac{C_n^2}{V(z)} - s\frac{\partial}{\partial z}(\frac{C_n}{\sqrt{V(z)}}) + i\frac{2EC_n}{\sqrt{V(z)}} \right \}$.
	\item $\frac{l_2^{''}}{l_2}$ =  $\left \{-{(-k_1k_    2 - sk_1^{'} - m_g^2) - h^{''}(z)} \right \} = \left \{-(E^2 - m_g^2) + \frac{C_n^2}{V(z)} + s\frac{\partial}{\partial z}(\frac{C_n}{\sqrt{V(z)}}) + i\frac{2EC_n}{\sqrt{V(z)}} \right \}$.
	\item From the above two expressions for $\frac{l_1^{''}}{l_1}$ and $\frac{l_2^{''}}{l_2}$, it can be observed that the factor that is different for  $\frac{l_1^{''}}{l_1}$ and $\frac{l_2^{''}}{l_2}$, (and thus eventually  $\frac{u_1^{''}}{u_1}$ and $\frac{u_2^{''}}{u_2}$), is the term $s\frac{\partial}{\partial z}(\frac{C_n}{\sqrt{V(z)}})$. This term causes the two spinor functions, $u_1$ and $u_2$, to have somewhat different derivatives, which implies that the Dirac spinor exists in a non-inertial frame. 
\end{itemize}
Let us then state the small gradient approximation as: 
\begin{eqnarray}
\nonumber \frac{\partial}{\partial z}(\frac{C_n}{\sqrt{V(z)}}) \sim 0, \\
\frac{\partial}{\partial z} \left \{-{(-k_1k_2 + sk_2^{'} - m_g^2) - h^{''}(z)} \right \} \sim 0.
\end{eqnarray}
In the small gradient approximation, one can define $q(z)$, to be the complex 3-momentum, where, 
\begin{eqnarray}
q(z)^2 = -{(-k_1k_2 + sk_2^{'} - m_g^2) - h^{''}(z)}\\
\approx \left \{ -(E^2 - m_g^2) + \frac{C_n^2}{V(z)} -  \frac{i2EC_n}{\sqrt{V(z)}} \right \}.
\end{eqnarray}
  The imaginary part of q(z), i.e. Imag(q(z)) gives the 3-momentum of the Dirac fermion, while the $Real(q(z)) + h^{-1}(z)$ determines the total decay rate. $Imag(q(z))$ is given by:
  \begin{equation}
  \label{eq:mom_obs}
	  Imag(q(z)) = \left ( \sqrt{\frac{1}{2} \left ( -Real(q^2) + |q^2|) \right )} \right ).
  \end{equation}
	   This could then be interpreted as the "instantaneous" 3-momentum. Alternatively, $\frac{1}{Imag(q(z))}$ can be seen as the "instantaneous" de-Broglie wavelength inside the thermal bath. 
  The change in the 3-momentum leads to a time lag as the fermion passes through the thermal bath, which can then be measured.
  We can then write the solution as:
\begin{equation}
\label{eq:smallgrad}
u_j(z) = h^{-1}(z)l_j(z) = e^{d(z)}e^{q_r(z)}e^{iq_i(z)},
\end{equation}
where $q_r(z)$ = Real(q(z)), $q_i(z)$ = Imag(q(z)),
and $d(z) = \frac{1}{4}\int \frac{V'(z)}{V(z)} dz$ (from Eq. ~\ref{eq:hz}).
  Thus, the solution is seen as a product of three terms, namely, 
  \begin{enumerate}[a)]
	  \item a scale invariant factor, $e^{d(z)}$, which depends only on the thermal gradient, $\frac{V'}{V}$, 
  \item a decay factor, $e^{q_r(z)}$, which depends on the temperature, and
  \item a phase factor, $e^{iq_i(z)}$.
	  \end{enumerate}
  It can also be inferred that, if the temperature gradient and the initial momentum of the fermion are in perpendicular directions, then the change in the momentum along the direction of the temperature gradient would lead to curving of the fermion path.

  The curving of the fermion path is reminiscent of the geodesic equations in Sec.~\ref{sec:geodesic}, albeit in a quantum mechanical context.
\subsubsection{Large temperature gradient}
In this case, $e^{(\int q(z)dz)}$ would not solve Eq.~\ref{eq:leq}. The solution has to be obtained numerically. 
Secondly, $\frac{l_2^{''}}{l_2} - \frac{l_1^{''}}{l_1} = 2s\frac{\partial }{\partial z} \frac{C_n}{\sqrt{V(z)}}$ and thus, $\frac{l_1^{''}}{l_1} \ne \frac{l_2^{''}}{l_2}$. For the above reasons, it is not straightforward to determine the 3-momentum observable. We now attempt to interpret the solution $l_1(z)$ and $l_2(z)$ obtained numerically. Define,
   		\begin{equation}
			l(z) = 
		\left( \begin{array}{c}
			l_1\zeta_0\\
			l_2\zeta_0\\
		\end{array} \right ), 
   		\end{equation}
			and
   		\begin{equation}
			\label{eq:spinorgen}
			u^{f}_p = 
		\left( \begin{array}{c}
			\sqrt(p.\sigma)\zeta_0\\
			\sqrt(p.\bar{\sigma})\zeta_0\\
		\end{array} \right ), 
   		\end{equation}
		with $\sigma$ being the Pauli spin matrices, and $\zeta_0$ was defined previously as part of Eq.~\ref{eq:spinor_u}.

		Fourier decompose $l(z)$ as:
\begin{equation}
\label{eq:fourier}
l(z) = \int c_p u^{f}_p e^{ipz} dp, 
\end{equation}
where,
		\begin{equation}
			\label{eq:invfourier}
c_p = \frac{1}{2m_g}\int u^{f\dagger}_p\gamma^0 l(z) e^{-ipz}dz.
		\end{equation}
		The decomposition allows the wave-function $l(z)$ to be interpreted  as a superposition of several fermionic waves of 3-momentum $p$, and amplitude $c_p$. 
\subsection{Numerical analysis and Basic Setup}
\label{sec:numbase}
   In this proposed setup, a fermionic particle, say an electron, is fired through a plasma and is detected on the other side of the plasma. Both the source and detector are in a near vacuum, i.e. close to zero temperature, and the high temperature plasma is sandwiched in-between.
\begin{figure}
\includegraphics[width = 80mm,height = 80mm]{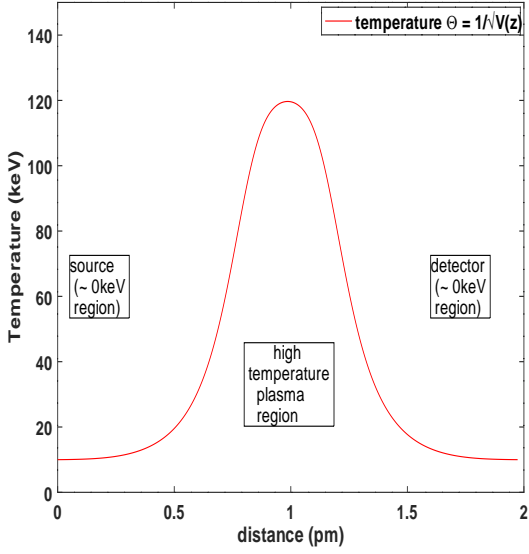}
	\caption{Plot of $\Theta(z)$ used for numerical analysis.}
\label{fig:setup}
\end{figure}
   Figure~\ref{fig:setup} shows the experimental setup. The temperature is Gaussian-like (but not Gaussian). The temperature, $\Theta$, and its derivative,  both approach 0 on either side, near the source and the detector.
   The wave equation near the source and the detector would then asymptotically tend to the vacuum solution of the Dirac equation.
The wave-function of the fermion emanating from the source would be of the form:
\begin{equation}
\psi = \sum_n a_n u_n(z)e^{-jh_{00}p^0_n\tau},
\end{equation}
where, $p^0_n = \frac{C_n}{V(z)}$, are the various Matsubara frequencies, with the constraint $\sum_n (a_n)^2 = 1$ and 
\begin{math}
	u_n(z) =  
		\left( \begin{array}{c}
			u_{1n}(z)\zeta_0\\
			u_{2n}(z)\zeta_0\\
		\end{array} \right ).
\end{math}
The Dirac equation is numerically solved for $n = 0$ and $n=20$, i.e., $C_0 = \frac{\pi}{\beta_0}$ and $C_{20} = \frac{41\pi}{\beta_0}$.
   The temperature curve, $\Theta$, depicted in Fig.~\ref{fig:setup}, is constructed using $V(z)$ as follows. Define $\frac{\partial V(z)}{\partial z}$ as:
   \begin{eqnarray}
	   \nonumber	   \frac{dV}{dz} = V_{init} \Bigg (  -\Big [\frac{1}{\sqrt{2\pi\sigma_1^2}} \exp(-\frac{(z - z_1)^2}{2\sigma_1^2}) \Big ] \\
	   +  \Big [\frac{1}{\sqrt{2\pi\sigma_2^2}} \exp(-\frac{(z - z_2)^2}{2\sigma_2^2} ) \Big ] \Bigg ),
   \end{eqnarray}
   Subsequently, $V(z)$ and $\Theta(z)$, are determined as:
   \begin{eqnarray}
\label{eq:vz}
	   V(z) = \int_{z0}^{zl} \frac{dV}{dz} dz,~~~~~
	   \Theta(z) = \frac{1}{\sqrt{V(z)}}.
   \end{eqnarray}
	   The temperature, $\Theta(z)$, for $\sigma_1 = \sigma_2 = 0.81$, and ($z_1, z_2$)  = ($0.39460$pm, $1.5784$pm), is depicted in Fig.~\ref{fig:setup}.
	   This construction ensures that both $V(z)$ and $\frac{dV}{dz}$ are nearly flat near the boundaries, allowing both $\Theta$ and curvature, $f(z)$, to be flat and small near the boundary.
	   The effective mass of the electron taken $=m =m_g - i\gamma^0E$, where $m_g = 0.511MeV$ and $E = 5MeV$, for simulation.

 A linearly spaced grid of fifty thousand points was chosen for the simulation, and ranged from $z=z0=0$ to $z=zl = 10MeV^{-1} \approx 1.973$pm. 
	The initial conditions for simulations were estimated as:
	$u_1^{sim}(zl) \approx u_1^{flat}(zl)$, and
	$u_1^{sim'}(zl) \approx u_1^{flat'}(zl)$ (refer Eq.~\ref{eq:flatsol}).

\begin{figure}
\includegraphics[width = 80mm,height = 80mm]{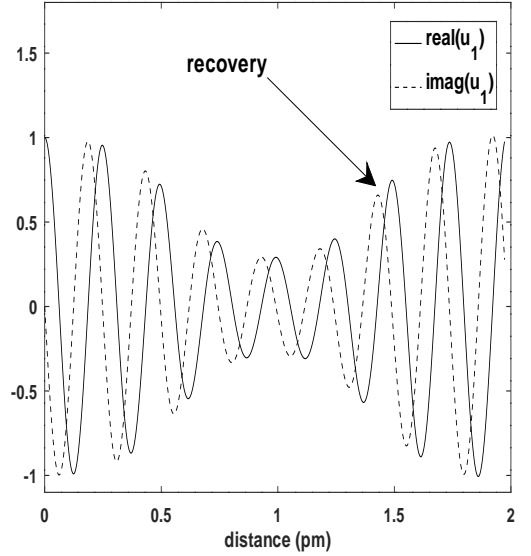}
	\caption{Plot of the wave-function $u_1(z)$ for Matsubara frequency $C_0 = \frac{\pi}{\beta_0}$.}
\label{fig:l1}
\end{figure}

\begin{figure}
\includegraphics[width = 80mm,height = 80mm]{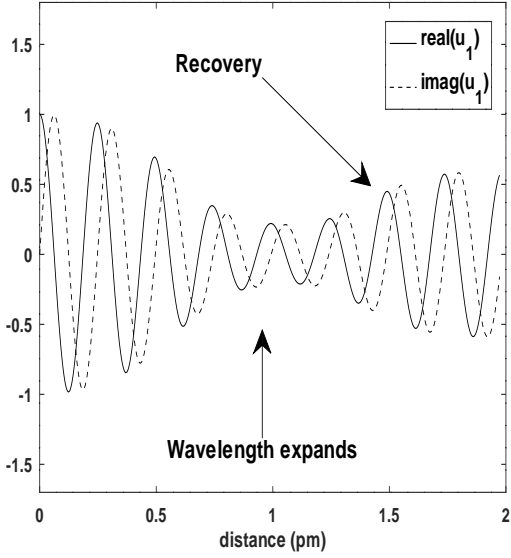}
	\caption{Plot of the wave-function $u_1(z)$ for Matsubara frequency $C_{20} = \frac{41\pi}{\beta_0}$ (Euclidean Signature).}
\label{fig:l1_20}
\end{figure}

\begin{figure}
\includegraphics[width = 80mm,height = 80mm]{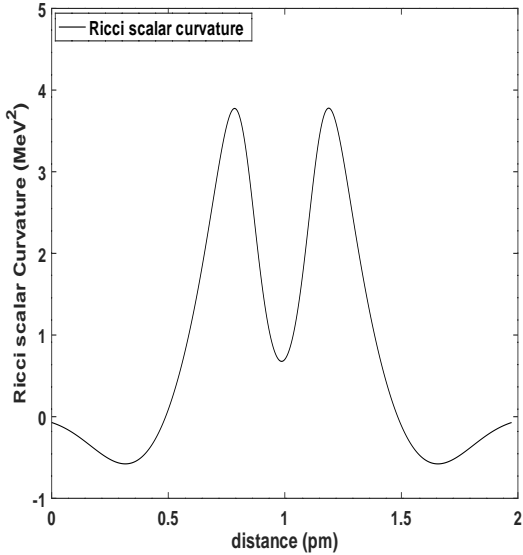}
	\caption{Plot of the Ricci scalar curvature at various values of $z$.}
\label{fig:ricci}
\end{figure}

	   We now analyze the general solution, which has been obtained numerically, for a large temperature gradient. Figures~\ref{fig:l1} and~\ref{fig:l1_20}, depict the waveforms, $Real(u_1)$ and $Imag(u_1)$, for the Matsubara frequencies, $C_0 = \frac{\pi}{\beta_0}$ and $C_{20}=\frac{41\pi}{\beta_0}$.
It can be seen in Fig.~\ref{fig:l1_20}, that in the region where the temperature decreases (refer corresponding temperature curve in Fig.~\ref{fig:setup}), the wave-function actually recovers, after a period of reduction in magnitude, due to the thermal bath. 
The reduction in magnitude may be interpreted as the partial loss of the particle wave-function to the thermal bath, and the recovery as the regain of the particle wave-function from the thermal bath.
The region where the recovery happens is actually a region with non-zero temperature. In any region of the thermal bath, comprising of non-zero temperature, one would expect a further reduction in the wave-function, and not an enhancement.
This enhancement is then purely due to the temperature gradient effect, as predicted by the proposed hypothesis (analogous to the $e^{d(z)}$ factor in Eq.~\ref{eq:smallgrad}).
The recovery in the wave-function would only be partial, as a result of decay due to the thermal bath 
(analogous to the $e^{q_r(z)}$ factor in Eq.~\ref{eq:smallgrad}).
	The Ricci scalar curvature, $R(z)$, corresponding to the temperature curve in Fig.~\ref{fig:setup} is plotted in Fig. \ref{fig:ricci}.
\subsection{Dirac fermion in gravity induced space-time curvature}
\label{sec:diracgravity}
\begin{figure}
\includegraphics[width = 80mm,height = 80mm]{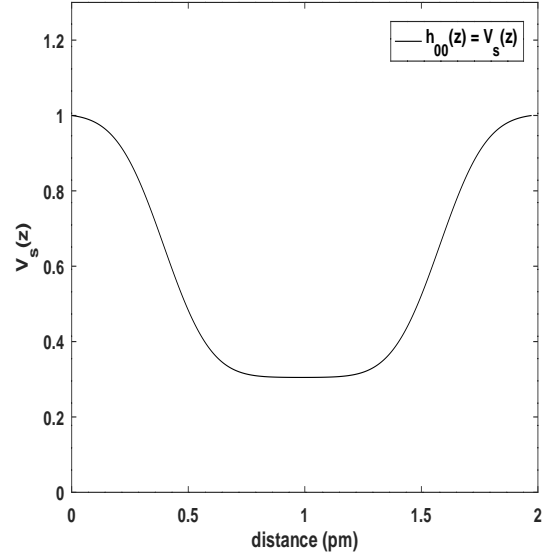}
	\caption{Plot of $g_{00}(z) = V_s(z)$.}
\label{fig:varr}
\end{figure}

\begin{figure}
\includegraphics[width = 80mm,height = 80mm]{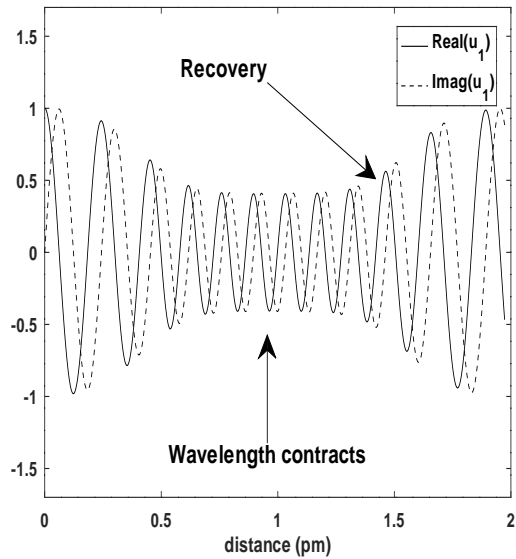}
	\caption{Plot of the wave-function $u_1(z)$ in gravity induced space-time curvature (Lorentzian Signature)}
\label{fig:u1dirac}
\end{figure}

Towards the purpose of comparison with the results obtained in Sec.~\ref{sec:numbase}, the Dirac fermion is now solved in gravity induced space-time curvature.
Let us consider the metric,
\begin{equation}
\label{eq:rtmetric}
	ds^2 = g_{00}(z)dt^2 - dx^2 - dy^2 - dz^2,
\end{equation}
with $g_{00}(z) = V_s(z)$.
The function $V_s(z)$ is a scaled and normalized version of $V(z)$, determined in Eq.~\ref{eq:vz}. The final function, $V_s(z)$, is depicted in Fig.~\ref{fig:varr}. 
After a simple algebraic manipulation, the Dirac equation boils down to:
\begin{eqnarray}
\label{eq:rtsecondorder}
\nonumber	u_1'' - \frac{f}{4\sqrt{V_s}}u_1' + (k_1 k_2 + isk_1' - m_g^2)u_1 = 0,\\
	u_2'' - \frac{f}{4\sqrt{V_s}}u_2' + (k_1 k_2 - isk_2' - m_g^2)u_2 = 0,
\end{eqnarray}
where
\begin{itemize}
\item $k_1 = \frac{E + if(z)s/8}{\sqrt{V_s(z)}}$; $k_2 = \frac{E - if(z)s/8}{\sqrt{V_s(z)}}$; $f(z) = \frac{2V'_s(z)}{\sqrt{V_s(z)}}$.
\end{itemize}
Comparing Eq.~\ref{eq:rtsecondorder} with Eq.~\ref{eq:secondorder}, the two equations are very similar, but with some important differences, which are now analyzed.
Figure \ref{fig:u1dirac} shows the solution of Eq.~\ref{eq:rtsecondorder}.
Comparing Fig.~\ref{fig:u1dirac} and Fig.~\ref{fig:l1_20}, one can observe some of the important similarities and differences.
\begin{itemize}
\item Differences
	\begin{enumerate}
	\item At the center, the wavelength contracts (as a result of blue shift) in the case of Fig.~\ref{fig:u1dirac} (Lorentzian signature), which is a well known result due to time contraction. However the wavelength marginally expands in case of Fig.~\ref{fig:l1_20} (Euclidean signature). The expansion in the case of the Euclidean signature, can be explained as slowing down of the fermion in a thermal medium having a higher temperature.
	\item When one compares the wave-function at the left and right end of the plot in Fig.~\ref{fig:l1_20} (i.e. at $0$pm and $2$pm), 
an overall decay of wave-function emerges in the case of the Euclidean signature.
			There is no overall decay for the Lorentzian signature (Fig.~\ref{fig:u1dirac}). The decay of a wave-function in a thermal medium is expected conventionally. 
	\end{enumerate}
\item Similarities
	\begin{enumerate}
	\item There is a sharp dip in the wave-function in both the figures between $0.5$pm and $1.3$pm. In the case of Fig.~\ref{fig:l1_20}, the dip is due to the much higher temperature in the region (refer Fig.~\ref{fig:setup}), leading to a higher decay. 
In the case of Fig.~\ref{fig:u1dirac}, the dip is due to space-time warping.
	\item Beyond roughly $1.3$pm, there is a recovery of the wave-function in both cases. In the case of Fig.~\ref{fig:u1dirac} (Lorentzian signature), it is easily explained by the un-warping of space-time. However, in the case of Fig.~\ref{fig:l1_20} (Euclidean signature), the recovery has no obvious explanation. 
Since it is a non-zero temperature everywhere, one would expect the wave-function to continuously decay. The magnitude of the decay may vary, but one would expect it to only decay, and never recover, as long as the temperature is non-zero. However, there is a recovery in regions of negative thermal gradient. 
			As mentioned earlier in Sec.~\ref{sec:numbase}, this seems to be purely the result of modeling thermal gradients as a curved Euclidean space, and probably, without any other obvious rationale. 
	\end{enumerate}
\end{itemize}
The question is, does this recovery actually happen? Thermal experiments can be performed due to the much lower energy involved. If the recovery of the wave-function in a thermal medium with negative thermal gradient is indeed observed, then it indicates that the curved Euclidean space is indeed the correct mathematical tool for modeling thermal gradients.

In case the recovery is not observed, then either some additional terms in the Lagrangian or Hamiltonian may be required, or in the worst case, the curved Euclidean space concept does not describe a thermal system with spatial gradients. In either case, it could lead to better understanding of the physics of thermal systems.


\section{Interpretation of the Matsubara Eigenstates, and other considerations}
\label{sec:interp}
\subsection{Interpretation of the Matsubara Eigenstates and energy E}
\label{sec:mats}
This section outlines a physical interpretation of the subtle aspects of the formalism developed so far.
Section~\ref{sec:scalarfield} looks at the bulk thermodynamic properties, while Sec.~\ref{sec:dirac} analyzes the properties of a single eigenstate of an external particle, i.e., a particle which is not part of the thermal medium. Both the scenarios are now compared.

For the purpose of the physical interpretation of aspects like the Matsubara frequency, it is not a necessity to consider a curved space, and hence we revert to the uniform temperature case, i.e., $h_{00}=1$.
In the original Matsubara formalism, the imaginary time theory does not introduce the particle energy, $E$, in the Lagrangian. 
The energy, $E$, is the energy of the particle when it emanates from the source. Clearly, $E$ is an intrinsic property of the particle, and not the result of interaction with the thermal medium. It does not obey Fermi-Dirac or Bose-Einstein distribution.
The introduction of $E$ has been enabled as a result of modeling the 8-D space, $\beta^{\mu} \times x^{\nu}$, since time axis is treated as a separate dimension from temperature.
The viewpoint of 8-D space $\beta^{\mu} \times x^{\nu}$, is a subtle difference having profound implications.
In Eq.~\ref{eq:nograd}, if we remove the contribution of $E$, by substituting $E=0$, we get,
        \begin{eqnarray}
        \label{eq:noE}
        \nonumber       u_1'' +  ( -\frac{C_n^2}{{V_0}}  - m_g^2)u_1 = 0,\\
        \nonumber       u_2'' +  ( -\frac{C_n^2}{{V_0}}  - m_g^2)u_2 = 0.\\
        \end{eqnarray}
The solution is straightforward, namely, $u_i = e^{\pm\sqrt{\frac{C_n^2}{{V_0}}  + m_g^2}z}$, where $i = 1,2$. This is a pure decay (enhancement) function, and the decay rate (enhancement rate) increases with mass. 
There is no real valued momentum observable (wave-function of the form $\exp(ipz)$), and the lack of the momentum observable is independent of the particle energy. This seems unrealistic. 
It is evident that at sufficiently high energy, a particle will have some momentum inside a plasma, and not just decay.

Let us take the example of bottom quark in a QGP. Taking $m_g  \approx 4.5GeV$, and radius of QGP $\approx 5 fm \approx 25GeV^{-1}$, we get $e^{-m_g z}\approx 2.9 \times 10^{-50}$. If this were correct, there would not be any bottom quark production at all in a QGP.  
However, there is sufficient production of these particles in QGP~\cite{hq1, hq2, hq3, hq4}. 
While, the interaction Lagrangian, in the imaginary time formalism, leads to reasonable results for the bulk properties like entropy etc., the Euler-Lagrange equations of motion arising from the free field Lagrangian, do not lead to meaningful results. 
The drawback can be root caused to the space $(i\beta,x,y,z)$. This space does not lend itself to a Lorentz invariant measure.
However, the 2-norm in the $\beta^{\mu}\times x^{\nu}$ space is Lorentz invariant, as shown in Eq.~\ref{eq:8D}. In a 8-D space, the consequence of the presence of the time axis is the introduction of particle energy, $E$, via the operator $\frac{i\partial}{\partial t}$. 
The presence of $E$ leads to the availability of a real valued momentum observable, as detailed in Sec.~\ref{sec:dirac}

To summarize, the introduction of $E$ leads to the following, highly desirable properties:
\begin{enumerate}
\item Enables modeling of the particle energy when it emanates from the source. This particle energy is distinct from the kinetics of the thermal medium.
\item Introduction of a momentum observable in the Euler Lagrange equations of motion.
\item In the limit of zero temperature ($\beta \rightarrow \infty$), one gets back the zero temperature Dirac equation.
\end{enumerate}

The Matsubara frequencies are viewed upon as the poles of the Bose-Einstein or Fermi-Dirac distribution. Consequently, the Matsubara frequency can be viewed as imaginary energy, $iE^c$ (= $iC_n/\sqrt(V)$), and is unrelated to the particle energy $E$. Thereupon, the energy, $E^c$, would be the energy of interaction of the particle with the thermal medium. 
The periodicity of the Euclidean space, effects the interaction energy, $E^c$, to be quantized. The interaction, then, leads to the decay of the wave-function, as depicted in Eq.~\ref{eq:flatsol}. This seems to be a reasonable outcome, as interaction may lead to a part of the particle wave-function merging with the thermal medium.
The various Matsubara frequency eigenstates, would then imply, that the particle is in a superposition of Matsubara frequency eigenstates, wherein, each eigenstate determines the degree (or energy) of interaction with the thermal medium.
Ergo, the spatial variation of temperature, would naturally lead to varying degrees of interaction of the particle with the thermal medium, giving way to the spatial variation of the Matsubara frequency. 

The standard imaginary time formalism can be seen to be a limiting case when $E\rightarrow 0$. In Sec.~\ref{sec:dirac}, an external particle of energy, $E$, has been analyzed. This particle had not equilibrated and was distinct from the thermal system. 
If, however, the particle "dissolves" or "equilibrates" and becomes a part of the thermal system, then, the only energy it would posses, would be the thermal interaction energy, $E^c$. It would no longer posses its own energy, $E$, which is distinct from the thermal system. 
The momentum of a single particle becomes meaningless, as only the ensemble average carries physical significance. The single particle Dirac equation seems to become redundant. 
Thus, the classical imaginary time formalism can be used to model the bulk properties of an ensemble of particles which have been equilibrated and have become part of the thermal system.
The concept of curved Euclidean space, obtained by recasting the temperature variation as a variation in the metric, is applicable in either case.
As a case in point, Sec.~\ref{sec:scalarfield} explores the equivalence in the context of bulk thermodynamic properties in a conventional 4-D Euclidean space, while, Sec.~\ref{sec:dirac} explores the external single particle behavior and includes $E$ in the modeling.

The application of the Dirac's theory, rather than the Field theory, leads to approximations, because the contributions due to particle creation and annihilation are ignored. However, this approximation should be much smaller for QED, relative to QCD, due to the much smaller QED coupling constant. 
The approximation may also work well for heavy quarks, such as the bottom or top quark, where the corrections due to the creation and annihilation of particles may be small.

\subsection{The Grand Canonical Ensemble}
\label{sec:gce}
The equivalence between the curved Euclidean space and spatial thermal variations can be extended to the grand canonical ensemble. This allows one to explore the effect of the chemical potential on the Matsubara frequency. 
In the case of uniform temperature, $\beta$, the partition function is, $Z = Tr\exp[\beta(H-\mu Q)]$, where, $\mu$ is the chemical potential, $H$ is the Hamiltonian  and $Q$ is the conserved charge with $Q = \int j^0 d^3 x$.
In the case of a thermal system with spatial thermal variations, the partition function with a time independent chemical potential, $\mu$, can be written as:
 \begin{eqnarray}
 \nonumber	Z = Tr \exp \left ( -\int \int_0^{\beta(\bx)} \left (\sH - \mu j^0 \right ) d\tau d^3x \right )\\
 	= Tr \exp \left ( -\int \int_0^{\beta_0} s(\bx) \left (\sH - \mu j^0 \right ) d\tau d^3x \right ),
 \end{eqnarray}
where $\sH$ is the Hamiltonian density. 
This suggests, that, like the case of the canonical ensemble, the grand canonical ensemble may also be recast in the framework of the curved Euclidean space, with the metric, $diag[s(\bx)^2, 1,1,1]$. 
Let us consider the following Lagrangian density for a fermion in the curved  Euclidean space given by the metric $diag[s(\bx)^2, 1,1,1]$.
\begin{equation}
	\sL = \bar{\psi} \left ( ie_a^0 \gamma^aD_{i\tau} + i\bf{\gamma.\nabla} - m \right ) \psi,
\end{equation}
where, $e^a_{\mu}$ is the vierbein, and the transformation, $t \rightarrow i\tau$ is incorporated in the spinor covariant derivative, $D_{i\tau}$.
The conserved charge density for the above Lagrangian density is 
\begin{equation}
	j^0 = \bar{\psi}(e_a^0 \gamma^a )\psi,
\end{equation}
With all this, the partition function can be written as:
 \begin{eqnarray}
 \nonumber	Z = \int D\psi^{\dagger} D\psi \exp \Bigg [ \int_0^{\beta_0} \int d^3x \sqrt{g_e}\\ 
 	\bar{\psi} \left ( ie^0_a\gamma^a D_{i\tau} + i\bf{\gamma.\nabla} - m + \mu e^0_a \gamma^a \right ) \psi \Bigg ], 
 \end{eqnarray}
 where, $\sqrt{g_e} = s(\bx)$ as before.
 If we decompose, $\psi(\bx,\tau) \rightarrow \sum_n \exp(-iC_n\tau) \psi_n(\bx)$,
 the term, $(ie^0_a\gamma^aD_{i\tau} + \mu e^0_a\gamma^a)\psi(\bx,\tau) \rightarrow \left [ e^0_a\gamma^a (-iC_n + \mu )\psi_n(\bx) + {\it spin~connection~terms} \right ]$. The shift in the Matsubara frequency, along the imaginary axis, due to the chemical potential, is akin to the uniform temperature case. 

In Sec.~\ref{sec:dirac}, the eigenstate transfiguration of a Dirac particle is modeled. In case the chemical potential of the thermal medium, were to affect the Dirac spinor, the results of Sec.~\ref{sec:dirac}, would be an accurate representation, predominantly for those thermal systems, where chemical potential is close to zero.

It is planned to probe the grand canonical ensemble case, in more detail, in future research.

\subsection{The 8-D space}
Another significant outcome of the 8-D point of view, is the doubling of the degrees of freedom.
The expansion to 8-D space from 4-D space involves the introduction of additional variables, namely, the 4-D conjugate momenta, $\omega_{\mu}$ in the $\beta^{\mu}$ subspace. In thermo field dynamics~\cite{tft1, tft2}, the doubling of degrees of freedom is shown to be a necessity, in order to define a thermal vacuum. 
The product space, $|p^{\mu}\rangle \otimes | \omega^{\nu} \rangle$, may seem to correspond to the tilde and non-tilde states, $|\tilde{m}\rangle \otimes |m\rangle$, in thermo field dynamics~\cite{tft1,tft2}, in the sense of the degrees of freedom being doubled. 
It remains to be seen, as to the extent the correspondence can be developed further.

Introduction of the 8-D, $\beta^{\mu} \times x^{\nu}$ space, leads to a field theory in a eight dimensional space.
There has been research in the renormalizability of $\phi^4$ theory in higher dimensional space (for example, Ref.~\cite{phi4}). Unfortunately, adaptation to thermal field theory may be a bigger challenge due to the presence of discrete frequencies.
However, Eqs.~\ref{eq:8D} and~\ref{eq:8Dmetric}, indicate that the 8-D space is not irreducible. It can be factored into two 4-D spaces. The factorization may likely enable analyzing the $\phi^4$ theory in dimensions lesser than 8-D.
   These issues are hoped to be addressed in future research.

\section{Conclusion} 
\label{sec:conclusion}
Starting with the help of a Polyakov loop correlator, we have propounded a mathematical equivalence between the temperature gradient and the curvature of the Euclidean space. The variation in the temperature is recast as a variation in the metric.
Bulk thermodynamic parameters, like the energy expectation value, Helmholtz free energy and entropy is calculated for a neutral scalar field in a thermal bath with spatial thermal gradients, using the concept of a curved Euclidean space. 
Subsequently, the Dirac equation for an external massive fermion is solved in the curved Euclidean space. 
    This analyzes the behavior of an individual eigenstate, instead of examining only bulk thermodynamic properties.
    The recovery of the fermion wave-function in the regions of negative thermal gradient is unexpected, and thus, can be a key indicator of the physical validity of the curved Euclidean space model.
 In case the recovery does not happen, it may likely require further research into discovering a valid model for systems with thermal gradients.
    The properties of the Dirac function, highlighted in this paper, is expected to carry over to a field theoretic approach as well, as the basis functions can remain the same.

    If the Euclidean space model, indeed models the quantum behavior in a thermal medium, then, it may have other implications.
The similarity of the particle behavior in the Euclidean and Lorentzian spaces, may lead to better understanding of particle behavior in a curved Lorentzian space-time. 
  As a case in point, for a scalar field, $\phi$, the curved Euclidean space experiments may help establish, whether, $R\phi^2$ is a required term in the Lagrangian.

  It would be desirable to generalize the proposed formalism to a covariant framework. A fully covariant field theoretic approach can help analyze the vacuum structure.
 

\acknowledgments 
I would like to thank Prof. Michael E. Peskin (SLAC National Accelerator Lab.) for useful discussions.

\end{document}